\documentclass[journal]{IEEEtran}

\usepackage{amsmath,amsfonts}
\usepackage{graphicx}
\usepackage{booktabs}
\usepackage{multirow}
\usepackage{array}
\usepackage{cite}
\usepackage{url}
\usepackage{balance}
\usepackage{enumitem}
\usepackage{siunitx}
\usepackage{subcaption}
\usepackage{tabularx}
\usepackage{adjustbox}
\usepackage{makecell}

\usepackage{comment}

\newcommand{\rakaah}{rakaah}

\title{Prayer-Gait-Auth: Smartphone IMU-based Behavioral Biometrics from Structured Islamic Prayer Movements}

\author{
\IEEEauthorblockN{Hassan Hizeh, Anwar B. Alshaibani, Muhammad Mahboob Ur Rahman, Tareq Y. Al-Naffouri}\\
\IEEEauthorblockA{
Computer, Electrical and Mathematical Sciences and Engineering Division (CEMSE),\\ 
King Abdullah University of Science and Technology, Thuwal 23955, Saudi Arabia.
\\
\{hassan.hizeh,muhammad.rahman\}@kaust.edu.sa  
}
\thanks{The research reported in this publication was supported by funding from King Abdullah University of Science and Technology (KAUST) - KAUST Center of Excellence for Smart Health (KCSH), under award number 5932.}
}

\begin{document}
\maketitle

\begin{abstract}
Islamic prayer is a structured movement activity that offers a distinctive
setting for behavioral biometrics: all participants execute the same high-level
action sequence, so identity must be inferred from differences in execution. We collected inertial data from 95 participants
using their own smartphones during nightly congregational Islamic prayer (Taraweeh). A
label-aware pipeline converts the long recordings (with 8+ rakaah) into structurally
complete two-\rakaah{} behavioral samples (prayer units). This allows us to design unit-level and
complete-session behavioral biometrics protocols. To address arbitrary smartphone orientation in worshippers'
pockets, we study two motion representations: rotation-invariant magnitudes
with gravity-relative acceleration components, and a metadata-aware
Qibla-referenced canonicalization that harmonizes platform conventions,
reconstructs device-to-world attitude, corrects Android magnetic north to
true north via WMM2025, and expresses acceleration and angular velocity in a
common Qibla-left-up frame. Under the unit-level protocol, the invariant and
Qibla-referenced representations reach learned pairwise Random Forest
AUC/EER of 0.9932/4.07\% and 0.9923/3.96\%; under complete-session holdout,
0.9700/7.62\% and 0.9618/7.81\%. Qibla-frame directional ablations show the
complete six-axis representation is strongest overall, with acceleration
retaining most learned-verification performance and vertical motion the
strongest single-axis cue. A Qibla-referenced SimCLR experiment further
yields participant-template AUC 0.9805--0.9817 and EER 5.81--6.24\% across
two unit-level runs. Complementary signal-, descriptor-, prayer-component ablations,
supervised-CNN, and self-supervised analyses show participant structure is
distributed across movement dynamics rather than concentrated in one signal
or posture. These results establish structured prayer movement as a
measurable cross-session behavioral biometric, while identifying device
heterogeneity and long-term persistence as key directions for future
validation.

%

%Under the unit-level protocol, the invariant and Qibla-referenced representations reach learned pairwise Random Forest AUC/EER of 0.9932/4.07\% and 0.9923/3.96\%, respectively; under complete-session holdout, the corresponding results are 0.9700/7.62\% and 0.9618/7.81\%. Qibla-frame directional ablations show that the complete six-axis representation is strongest overall, with acceleration retaining most learned-verification performance and vertical motion providing the strongest single-axis cue. A Qibla-referenced SimCLR experiment further yields participant-template AUC 0.9805--0.9817 and EER 5.81--6.24\% across two unit-level runs. 

\end{abstract}


\section{Introduction}
Biometric recognition uses physiological or behavioral characteristics to
establish or verify identity. Behavioral traits are attractive for unobtrusive
and potentially continuous authentication, but their utility depends on a
balance between inter-person distinctiveness and intra-person repeatability
\cite{jain2004introduction,jain2007biometric}. This balance is particularly
challenging for motion biometrics because the observed signal can vary with
pace, physical state of the person, device placement, acquisition context, and sensor
characteristics.

This work aims to study behavioral/motion biometrics in the context of Islamic prayer (\emph{Salah} or \emph{Salat}), one of the most widely and regularly performed structured movement sequences in the world. It is observed by an estimated two billion Muslims worldwide, prescribed five times daily, and follows an essentially fixed ordered sequence of postures regardless of where or by whom it is performed. This combination of scale, regularity, and prescribed structure makes prayer movement an unusually well-powered and practically relevant setting in which to ask whether identity can be recovered from how a shared action is executed, and it opens a natural pathway from a behavioral-biometrics research question to real-world smartphone and wearable-sensing deployments used by a very large population.

On the sensing front, commodity smartphones that consist of an inertial measurement unit (IMU), have gained popularity. A variety of android and iOS apps allow capture of raw IMU data streams from accelerometers, gyroscopes, gravity sensors, along with
orientation information. The smartphone-based IMU data
have supported gait recognition, continuous authentication, grasp and hand
movement analysis, and related forms of behavioral identity inference
\cite{derawi2010unobtrusive,sitova2015hmog,sprager2015inertial,
alzubaidi2016authentication,gadaleta2016idnet,zou2018gait}. Most prior motion
biometrics, however, are observed during activities whose content or context can
vary across users. A prescribed movement sequence offers a complementary
setting: when participants perform approximately the same ordered actions,
identity must be expressed in \emph{how} those actions are executed, e.g., through
movement intensity, transition shape, temporal rhythm, postural control, and
pause duration (rather than through activity choice).

\begin{figure*}[htbp]
\centering
\includegraphics[width=0.90\textwidth]{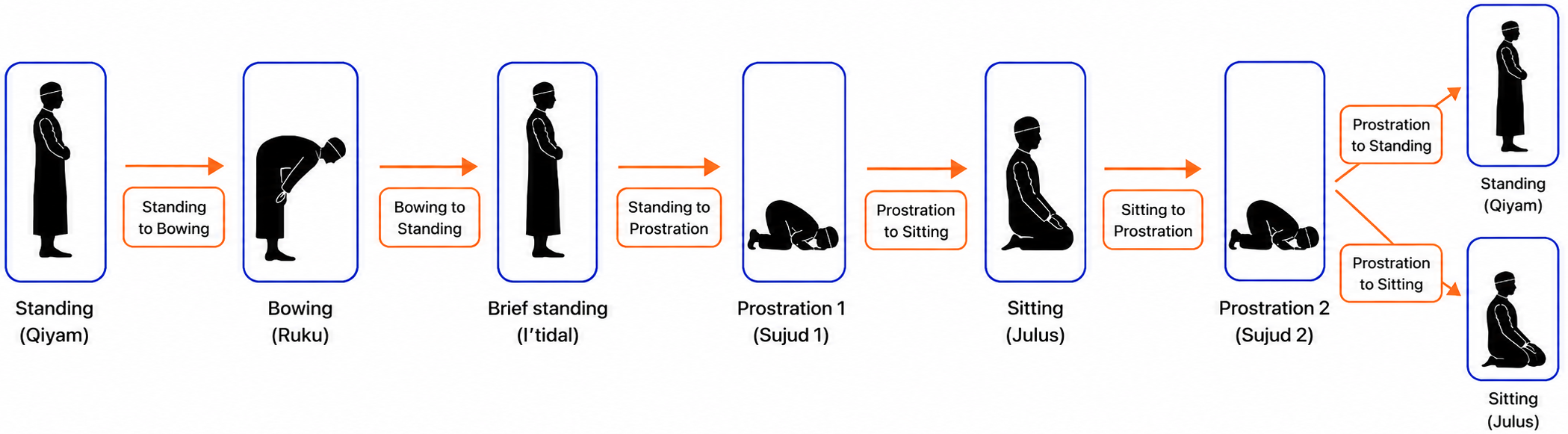}
\caption{Postures and transitions forming one rakaah.
The prescribed sequence is shared across participants; biometric analysis
therefore targets differences in execution rather than differences in activity
order. Note that as per Islamic convention, two consecutive rakaahs are paired together to constitute one complete two-\rakaah{} unit. }
\label{fig:prayer_structure}
\end{figure*}

Islamic prayer is a natural instance of such a structured behavior. A
\emph{Rakaah} makes a building block of Islamic prayer, and consists of an ordered sequence of the following postures/vocabulary: standing (\emph{Qiyam}), bowing (\emph{Ruku}), prostration (\emph{Sujud}), sitting (\emph{Julus}), and transitions between them (see Fig. \ref{fig:prayer_structure}). Existing prayer-sensing research has primarily addressed problems such as recognition of postures and transitions between them, deviations from regular prayer sequence due to forgetfulness, or rakaah counting
\cite{alghannam2016prayer,ali2018salat,ahmad2019sarm,koubaa2020salat,
vurgun2024salat}. The biometric question that this work studies is
fundamentally different: whether participant-specific execution is sufficiently
distinctive and repeatable to support identification and verification,
particularly when enrollment and query samples are acquired in different
recording sessions.

%\emph{Taraweeh} is a congregational nightly prayer held during Islamic month of Ramadan that provides unique opportunity to collect repeated observations of the prayer structure within one long continuous recording. 

In this work, IMU data were acquired from 95 participants who recorded themselves via their own phones placed in either of the two front pockets of their trousers, while they performed \emph{Taraweeh}, the nightly congregational prayer during the Islamic month of \emph{Ramadan}. We make the following definitions. i) A {prayer unit} (or a \emph{unit}) is a collection of two consecutive rakaah. ii) A single measurement session (or a \emph{session}) covers a single-night Taraweeh prayer, and consists of four or more consecutive units. Many participants contributed to multiple measurement sessions, corresponding to multiple nights. This allows us to do biometrics analysis at the unit (session) level by considering a unit (a session) as one behavioral sample. 

%In addition, we do a complete-session holdout, in which every unit from the query recording is excluded from preprocessing, training, enrollment, template construction, and pair generation. 

One challenge during prayer IMU data acquisition is the arbitrary orientation of the phone in the pocket of worshippers. This could erroneously rotate the same physical motion of the various concurrent participants of the congregational prayer into different arbitrary directions.
We study two complementary solutions to address this confound: i) a rotation-invariant magnitude+gravity-relative representation, and
ii) a metadata-aware directional representation that maps acceleration and angular
velocity into a common physical frame referenced to true north, vertical, and
the \emph{Qibla} direction\footnote{\emph{Qibla} or \emph{Kaa'ba} refers to the Grand mosque in Makkah, Saudi Arabia, which is considered as the holiest site by Muslims, and thus, Muslims throughout the world face towards Qibla direction ($21.4225^\circ\,\text{N}, 39.8262^\circ\,\text{E}
$) when it is time to pray, alone or with congregation.}.

\textbf{Contributions.}
This work makes the following principal contributions:
\begin{itemize}[leftmargin=*]
    \item We formulate structured congregational prayer as a behavioral
    biometric task and study 95 participants using real-world smartphone IMU
    recordings. 
    %segmented into quality-controlled two-rakaah behavioral units.
    \item We introduce and evaluate a metadata-aware Qibla-referenced
    canonicalization pipeline and compare it, on an exactly matched evaluation
    design, with a rotation-invariant magnitude+gravity-relative
    representation.
    \item We establish complementary unit-level and session-level protocols
    spanning nearest-neighbor identification, ML model-based classification,
    fixed-score pairwise verification, ML model-based pairwise verification, and
    participant-template matching.
    \item We characterize where participant-discriminative information appears,
    through signal-, descriptor-, prayer-component, and Qibla-direction
    ablations. {We also report supervised (CNN) and self-supervised (SimCLR/BYOL/DINO) analyses
    for rotation-invariant and Qibla-referenced IMU representations.}
\end{itemize}

\textit{Remark:}
We note that heterogeneous sensing hardware used by participants is a potential confound. Consumer smartphone MEMS
sensors are not metrologically identical: bias, scale-factor error, noise, and
platform-level processing can vary across handset models \cite{kos2016crossplatform,capuano2023smartphone}. Because the present collection intentionally used participant-carried phones, device identity can be correlated with participant identity. 
%Complete-session holdout eliminates recording overlap but is not equivalent to device-disjoint evaluation. 
We therefore treat handset effects as a potential confound rather than silently attributing all discrimination to behavior.

\section{Related Work}

The primary related work for this study constitutes the works that do motion analytics on prayer IMU data. The secondary (partially relevant) related work includes works that utilize inertial sensors for various kinds of behavioral/motion biometrics, e.g., gait analysis. 
This section summarizes selected related works from both literature domains. 

%\subsection{Inertial Behavioral Biometrics}

%Smartphone and wearable inertial sensors have been widely used for human activity recognition, typically by segmenting accelerometer and gyroscope streams and extracting temporal and spectral descriptors \cite{kwapisz2011activity,anguita2013public,reyesortiz2013ucihar,shoaib2014fusion,shoaib2015survey}. Deep temporal models increasingly learn such representations directly from sensor sequences \cite{zhang2022deep}. Activity recognition and behavioral biometrics nevertheless optimize different objectives: the former asks which activity occurred, whereas the latter asks which person generated a nominally comparable activity.

%Motion biometrics have been studied most extensively through gait and natural smartphone interaction. Inertial gait recognition captures participant-specific cadence, impact, and movement structure \cite{derawi2010unobtrusive,sprager2015inertial}; convolutional models further demonstrate identity learning from smartphone gait acquired in less constrained settings \cite{gadaleta2016idnet,zou2018gait}. HMOG showed that hand movement, orientation, and grasp dynamics accompanying ordinary smartphone use also support continuous authentication \cite{sitova2015hmog}. 

\subsection{Prayer Movement Sensing}

Existing work on prayer movement sensing has predominantly focused on activity recognition rather than identity authentication \cite{alghannam2016prayer,ali2018salat,ahmad2019sarm,koubaa2020salat,
vurgun2024salat}, using smartphone and smartwatch IMUs. Early efforts established foundational posture classification from accelerometer streams. Alghannam \emph{et al.}~\cite{alghannam2016prayer} analyzed 118 acceleration clips from six worshippers using a phone mounted on the upper back; their WEKA-based classifiers achieved 91.3\% ten-fold cross-validated accuracy. Alobaid \emph{et al.}~\cite{alobaid2018prayer} acquired a 20-participant hip-mounted smartphone dataset, designed a two-tier classifier that routed ambiguous posture cases to a dedicated second stage, achieving an accuracy of 93\%. 
%Ahmad \emph{et al.}~\cite{ahmad2019sarm} extended this line by modeling ordered Salah activities with SARM, while Koubaa \emph{et al.}~\cite{koubaa2020salat} applied deep learning to richer prayer sequences.

More recent work has reported near-saturated posture classification accuracy under increasingly constrained protocols. Syed \emph{et al.}~\cite{Syed2025Stacked} reported 100\% accuracy with stacked LSTM and BiLSTM networks, yet trained on only seven subjects and tested on three, a design that limits generalizability. Boudakkou \emph{et al.}~\cite{boudakkou2024leveraging} attained a mean average precision of 0.995 for posture recognition, but their system relies on smartphone camera imagery, which is unsuitable for private worship settings. Jahan \emph{et al.}~\cite{jahan2023smartwatch} collected a 30-subject wrist-worn corpus; under leave-one-subject-out evaluation, conventional classifiers averaged 77–84\% accuracy and fell below 70\% for some individuals. Subsequently, they utilized hand-coded semantic rules fused with dynamic time warping that boosted performance to 99.3\%.
%this relied on a template database constructed from \emph{all} enrolled subjects, risking evaluation leakage.

Beyond posture recognition, several studies address prayer within broader human activity recognition (HAR) frameworks. Vurgun \emph{et al.}~\cite{vurgun2024salat} treated prayer as one of eight activity classes using smartwatch sensors on fifty male participants, reporting 96.4\% multi-class accuracy but without decomposing prayer into constituent postures or transitions. Alselwi \emph{et al.}~\cite{al2022neural} tackled a different problem whereby they distinguish the five daily prayers by rakaah count from 134 smartwatch recordings, and achieve 87\% accuracy. Jarrah \emph{et al.}~\cite{Jarrah2025Elderly} employed dual waist-and-thigh IMUs to record 16 activities of daily living (including prayer) from 21 participants, yet their primary objective remained general HAR rather than prayer-specific posture segmentation. 
%Ali \emph{et al.}~\cite{ali2018salat} and others have similarly used smartphone accelerometers for prayer activity and posture recognition, while smartwatch IMUs have been evaluated for Salat recognition \cite{vurgun2024salat}. 
Recent work further explores contactless mm-wave radar for prayer tracking \cite{saifullin2026prayertracker} and smartphone IMU prayer data-based postural stability analysis \cite{anwaraaai27}.

%In those problems, the common movement structure is the target signal. In the present biometric task, that structure is controlled and the desired information lies in residual differences in execution.

\subsection{Inertial Behavioral Biometrics}

Behavioral biometrics exploits distinctive patterns in human motion, e.g., gait, typing rhythm, gestures, for authentication without explicit user cooperation~\cite{derawi2010gait}. The proliferation of IMUs in consumer wearables has made smartphones and smartwatches ubiquitous platforms for motion-based identity verification. Early foundational work by Gafurov~\emph{et al.}~\cite{gafurov2006biometric} demonstrated that waist-mounted accelerometers capture person-specific spectral features, while Makihara~\emph{et al.}~\cite{makihara2006silhouette} established silhouette-based gait recognition as a complementary vision modality. Derawi~\cite{derawi2010gait} provided a comprehensive survey of sensor types, feature families, and matching strategies, and Nickel~\emph{et al.}~\cite{nickel2011hmm} introduced hidden Markov models that leverage temporal stride sequencing to improve equal-error rates.

The integration of tri-axial IMUs into smartphones enabled ubiquitous gait authentication. Ngo~\emph{et al.}~\cite{ngo2015similar} proposed a similarity-based framework robust to walking-speed variation, while Thang~\emph{et al.}~\cite{nishida2014gait} achieved over 90\% authentication accuracy using frequency-domain features from pocket-carried devices. Muaaz and Mayrhofer~\cite{zhang2018smartphone} developed and tested their gait-based authentication system whereby participants carried phones at multiple body locations. Lee~\emph{et al.}~\cite{zhang2019gaitprint} explicitly evaluated cross-session gait authentication over multiple days, using sensor-coordinate compensation to reduce inter-session variability. Giorgi~\emph{et al.}~\cite{mondal2017continuous} extended this to continuous authentication, reducing the need for repeated explicit logins. Recent advances include GaitAuth, a CNN-BiLSTM pipeline with on-device processing achieving 94.3\% accuracy~\cite{vyshak2025gait}, and AMDFAuth, a deep reinforcement learning framework reaching 98.52\% accuracy with 0.94\% equal-error rate~\cite{li2025drl}. Zero-shot learning has been identified as a promising direction to reduce labeled-data requirements~\cite{abuhamad2020sensor}.

Wrist-worn IMUs present distinct challenges due to differing arm-swing dynamics and varying sensor orientation. Cola~\emph{et al.}~\cite{kuptametee2017smartwatch} showed that wrist-gait features, though noisier, achieve acceptable equal-error rate when aggregated over multiple gait cycles. Al-Naffakh~\emph{et al.}~\cite{alshehri2020smartwatch} evaluated smartwatch-based gait recognition from accelerometer and gyroscope data on 60 users under same-day and cross-day conditions, and later extended this line of work to continuous authentication using smartwatch motion data collected over multiple days~\cite{alshehri2021continuous}. The field has shifted from handcrafted features to end-to-end deep learning: Dehzangi~\emph{et al.}~\cite{chen2019gait} applied CNNs to time--frequency representations of IMU gait cycles, with multi-sensor fusion further improving identification performance, while Tran~\emph{et al.}~\cite{rida2018gait} benchmarked CNN and LSTM architectures, concluding that temporal recurrence is critical for capturing stride periodicity. Cao~\emph{et al.}~\cite{zhang2020deep} combined CNN and LSTM components with an attention mechanism and used orientation transformation to reduce sensitivity to sensor direction. 

%Despite progress, open challenges remain, including sensor heterogeneity, template aging, privacy concerns, and underexplored anti-spoofing mechanisms.

\subsection{Summary and Positioning}
Relative to this literature, this work differs along three axes. 
First, unlike existing prayer-sensing work, the
objective is participant identity rather than posture or \rakaah{} recognition.
Second, unlike gait and natural-interaction
biometrics, the target activity (Islamic prayer) is fully prescribed and repeated, so
activity-content variation is controlled by design rather than by
post-hoc filtering. 
Third, unlike most inertial-biometrics studies, we evaluate two coordinate
treatments with different invariance assumptions under an exactly matched
protocol. In addition, we separate a broad unit-level analysis from a strict
complete-session holdout evaluation so that distinctiveness and cross-session persistence
are not conflated. This combination is, to our knowledge, not addressed by
prior work in either the prayer-sensing or the inertial-biometrics literature.

%Prayer differs in that it is a long, semantically ordered, multi-posture sequence repeated by all participants, thereby reducing activity-content variation and emphasizing individual execution.

\section{The Dataset}
\label{sec:dataset}

%\subsection{Prayer Structure and Behavioral Unit}
%Table~\ref{tab:terminology} defines the movement labels used throughout the study, and Fig.~\ref{fig:prayer_structure} illustrates their ordering. One \rakaah{} comprises standing, bowing, two prostrations separated by sitting, and the associated transitions. A complete two-\rakaah{} unit contains this ordered sequence twice and serves as the principal behavioral sample.

\begin{comment}

\begin{table}[!t]
\centering
\caption{Prayer movement terminology used in the study.}
\label{tab:terminology}
\resizebox{\columnwidth}{!}{%
\begin{tabular}{ll}
\toprule
Term / label & Meaning \\
\midrule
Standing / qiyam & Upright standing posture \\
Ruku & Bowing posture \\
Ruku--Qiyam & Transition between standing and bowing \\
Sujud 1 & First prostration \\
Julus & Sitting between or after prostrations \\
Sujud 2 & Second prostration \\
Sujud--Qiyam & Transition from prostration to standing \\
Two-\rakaah{} unit & Two complete ordered \rakaah{} sequences \\
\bottomrule
\end{tabular}}
\end{table}

\end{comment}

%\subsection{Collection Protocol, Ethics, and Participants}

Data were collected during nightly congregational Taraweeh prayer while participants
carried their own smartphones in either of the two front pockets of their trousers\footnote{This research study was approved by the Institutional Biosafety and Bioethics Committee (IBEC) of the King Abdullah University of Science and Technology, Saudi Arabia, Protocol number: 24IBEC031 (date of approval: Jan. 4, 2026). All subjects provided their informed written consent before data collection. Data collection was carried out according to the Declaration of Helsinki.}. Acceleration, gyroscope, gravity, and orientation
streams were recorded at 100 Hz, using \emph{Sensor Logger} app. Recall that the nightly congregational Taraweeh prayer consists of at least 8 rakaahs (i.e., at least 4 prayer units).

The final cohort comprised 95 participants (75 male and 20 female), aged
19--56 years with mean age $29.9\pm8.9$ years. Participant-carried smartphones
were used intentionally, yielding heterogeneous consumer hardware rather than a
single calibrated reference IMU device. This improves ecological validity but it also implies
that device-dependent bias, scale factor, noise, and operating-system sensor
processing are not experimentally disentangled from participant identity. No
study-specific cross-device metrological calibration was performed.
Recordings were excluded when required signal streams were unavailable, when
the expected prayer structure could not be recovered, or when the prayer was
performed outside the congregational setting considered here.

\begin{figure*}[!t]
    \centering
    
    % Top Subfigure
    \begin{subfigure}{0.8\textwidth} % Adjust width as needed
        \centering
        \includegraphics[width=\textwidth]{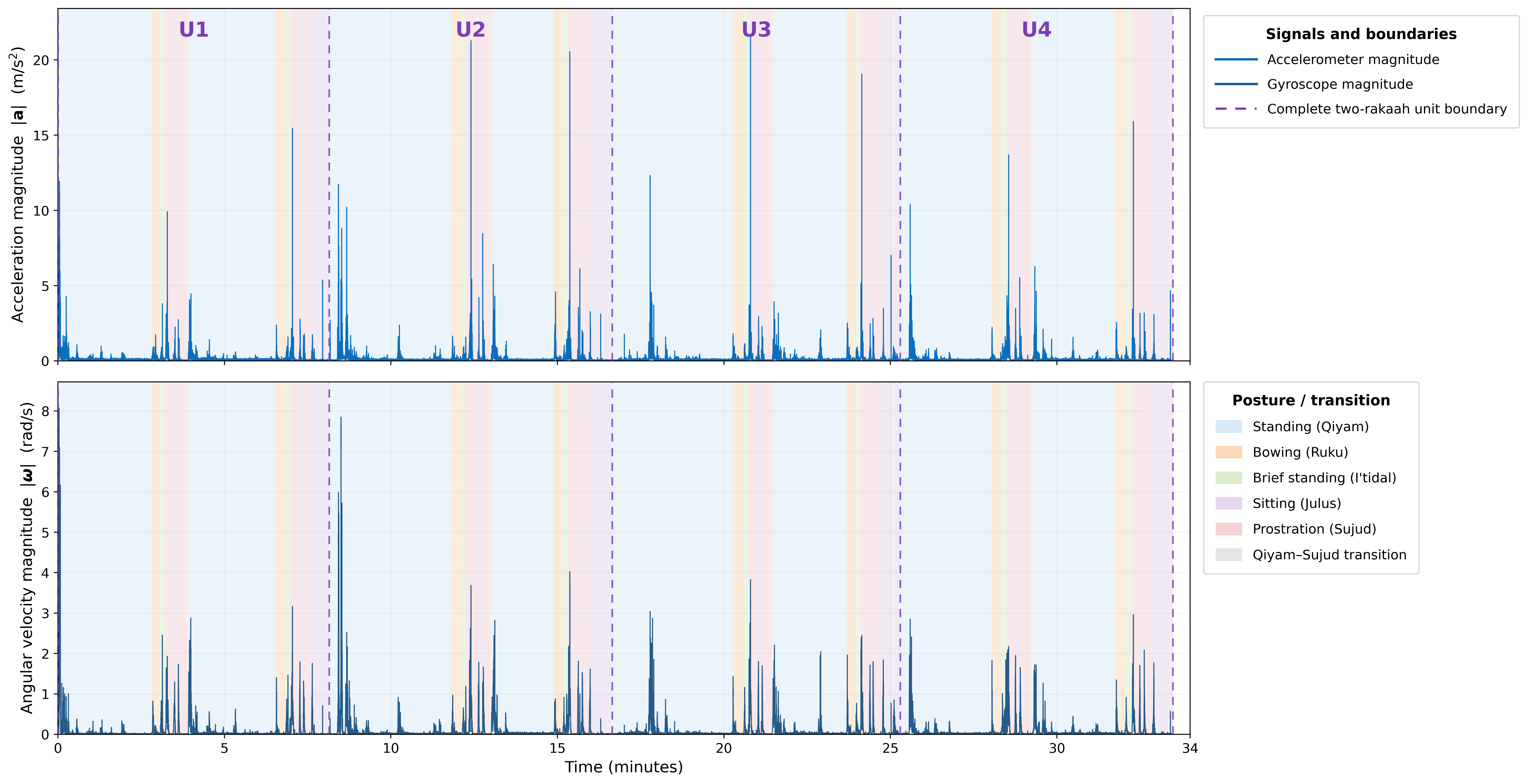} % Replace with your image file
        \caption{Full-session recording (a nightly Taraweeh prayer with 4 prayer units $U_1-U_4$ identified).}
        \label{fig:top_sub}
    \end{subfigure}
    
    \vspace{0.5cm} % Adds vertical space between the upper and lower subfigures
    % Leaving a blank line here forces the vertical stacking
    
    % Bottom Subfigure
    \begin{subfigure}{0.8\textwidth}
        \centering
        \includegraphics[width=\textwidth]{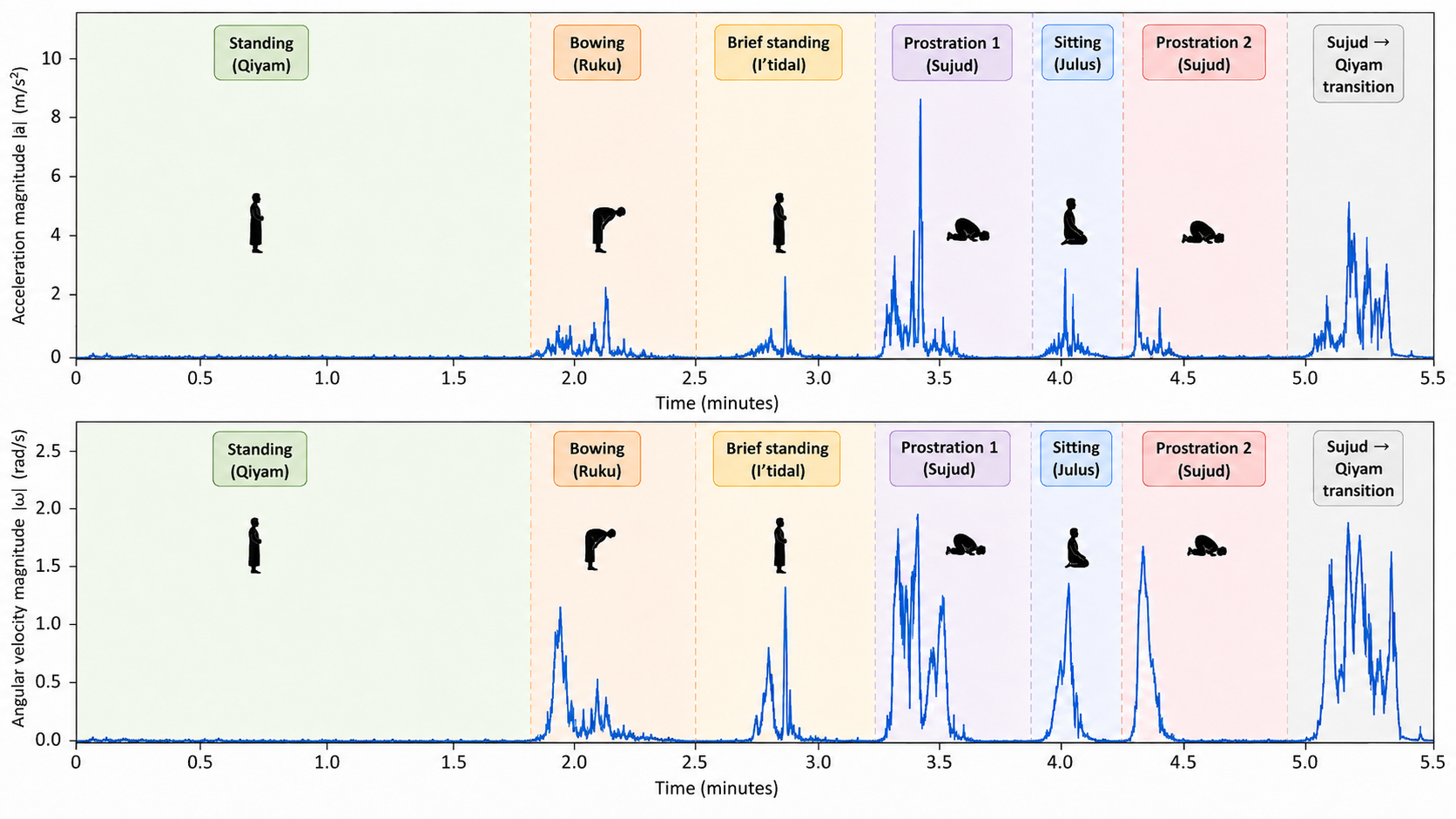} % Replace with your image file
        \caption{Single-rakaah recording (zoomed in from full-session recording).}
        \label{fig:bottom_sub}
    \end{subfigure}

    % Main Figure Caption and Label
    \caption{(a) full-session recording with unit boundaries identified, and (b) zoomed-in single-rakaah recording with various postures labeled, of an anonymous participant. }
\label{fig:segmentation}
\end{figure*}

\textit{Data Segmentation into Units:}
The processing pipeline identified 203 recording sessions, of which 196
contained the required streams and entered label-aware segmentation into units. The unit
boundaries were inferred from the ordered activity sequence while distinguishing
the sitting interval internal to a \rakaah{} from the sitting interval that
closes a unit. A unit was retained as complete only when the
expected one-\rakaah{} posture sequence occurred twice in valid order, including
two bowing events and two complete prostration cycles. Partial starts, partial
ends, and structurally incomplete unit candidates were excluded from complete-unit
analyses. A 10-s boundary trim was applied at each unit edge before feature
construction to reduce boundary uncertainty and phone-handling contamination.
Figure~\ref{fig:segmentation} shows an anonymized labeled session recording (along with unit-level labels/segments).

The segmentation stage detected 910 candidate units. Structural quality control
retained 785 complete units, and feature-validity checks yielded 780 units from
195 recordings (corresponding to 95 participants) for the principal analyses. 

\textit{Multi-Session Measurements:}
Complete-session
holdout requires the query participant to retain at least one independent
enrollment recording. A total of 32 participants meet this criterion, providing
132 held-out recordings and 527 query units. Single-session participants cannot
serve as cross-session query identities but remain, where applicable, as
competing enrolled identities or impostor references. Table~\ref{tab:dataset_summary}
summarizes the analysis populations.

\begin{table*}[!t]
\centering
\caption{Dataset construction and principal evaluation populations.}
\label{tab:dataset_summary}
\resizebox{\textwidth}{!}{%
\begin{tabular}{lcccl}
\toprule
\textbf{Stage or analysis} & \textbf{Participants} & \textbf{Recordings} &
\textbf{Units or queries} & \textbf{Role in the study} \\
\midrule
Discovered recordings & 95 & 203 & -- & Raw recording inventory \\
Signal-complete recordings & 95 & 196 & -- & Input to label-aware segmentation \\
Detected two-\rakaah{} candidates & 95 & 196 & 910 candidate units & Candidate unit pool \\
Structurally complete units & 95 & -- & 785 complete units & Quality-controlled behavioral samples \\
Handcrafted unit analysis & 95 & 195 & 780 units & Unit-level analysis and ablations \\
Complete-session holdout & 32 query-eligible & 132 held-out recordings & 527 query units & Cross-session evaluation \\
\bottomrule
\end{tabular}%
}
\end{table*}

% The asterisk (*) makes the figure environment span both columns

\begin{comment}
    
\begin{figure*}[!t]
\centering
\includegraphics[width=0.98\textwidth]{figures/segmentation_example_anonymized.png}
\caption{An example single-session recording of an anonymous participant that is labeled at posture level, along with detected complete two-\rakaah{} units.
The ordered activity sequence determines candidate unit boundaries and structural
quality control.}
\label{fig:segmentation}
\end{figure*}

\end{comment}

\section{Data Pre-processing: Coordinate Harmonization and Motion Representations}
\label{sec:representations}

%\textit{Coordinate Frames and Heterogeneous Smartphone IMUs:}
Phone-fixed axes vary with handset orientation, so identical physical motion can
appear under different coordinate rotations. Thus, we consider two compensation strategies: i) construct rotationally invariant or gravity-relative IMU representation, and ii) explicitly estimate attitude and rotate motion into a common
physical reference frame. The first removes orientation nuisance variables
analytically at the cost of directional detail; the second retains direction
but depends on orientation conventions and heading reference.

\textit{Cross-Platform Sensor Convention Harmonization:}
The smartphone prayer IMU data were recorded with Sensor Logger app. Its documentation specifies
platform-dependent (iOS vs. Android) acceleration/gravity and orientation conventions
\cite{sensorlogger_coordinates,sensorlogger_crossplatform,
sensorlogger_orientation}. 
The metadata associated with each recording is therefore used as the
authoritative source for standardization.
Let $\mathbf a_D(t)$, $\boldsymbol\omega_D(t)$, and $\mathbf g_D(t)$ denote three-axis linear
acceleration, angular velocity, and gravity expressed in the device-fixed frame
$D\in\mathbb R^3$.
%after (android/iOS) platform convention harmonization.
Acceleration and gravity are first mapped to a common device-axis convention,
\begin{equation}
\tilde{\mathbf a}_D(t)=s_p\mathbf a_D(t),\qquad
\tilde{\mathbf g}_D(t)=s_p\mathbf g_D(t),
\label{eq:platform_sign}
\end{equation}
where $s_p=-1$ for native non-standardized iOS recordings and $s_p=+1$ for
standardized iOS and Android recordings. No analogous global sign inversion is
applied to the gyroscope. This operation harmonizes documented platform
conventions. 
%it is not selected from biometric performance.

%Orientation samples are synchronized to native sensor timestamps using spherical linear interpolation (SLERP), which interpolates on the unit-quaternion sphere \cite{shoemake1985slerp}.

\color{black}

\subsection{Rotation-Invariant \& Gravity-Relative IMU Representation}
This method aims to remove arbitrary coordinate rotations analytically while
retaining movement intensity and gravity-relative motion structure. 
%Let $\mathbf a(t)$, $\boldsymbol\omega(t)$, and $\mathbf g(t)$ denote three-axis acceleration, angular velocity, and gravity after platform convention harmonization. 
The three rotation-invariant magnitudes are
\begin{equation}
 a_{\mathrm{mag}}=\|\tilde{\mathbf a}_D\|_2,\quad
 \omega_{\mathrm{mag}}=\|\boldsymbol\omega_D\|_2,\quad
 g_{\mathrm{mag}}=\|\tilde{\mathbf g}_D\|_2.
\end{equation}
With $\hat{\mathbf g}=\tilde{\mathbf g}_D/g_{\mathrm{mag}}$, acceleration is decomposed as
\begin{equation}
 a_{\mathrm{vert}}=\tilde{\mathbf a}_D^{\top}\hat{\mathbf g}_D,\qquad
 a_{\mathrm{horiz}}=\|\tilde{\mathbf a}_D-a_{\mathrm{vert}}\hat{\mathbf g}_D\|_2.
\label{eq:invariant_representation}
\end{equation}
The five signals
$[a_{\mathrm{mag}},\omega_{\mathrm{mag}},g_{\mathrm{mag}},a_{\mathrm{vert}},
a_{\mathrm{horiz}}]$ therefore provide analytical rotational invariance. 
This is because for any common proper rotation $\mathbf{Q}\in SO(3)$, the three vector norms, the gravity-relative
inner product, and the horizontal residual norm are all preserved. 

This representation requires no orientation metadata and serves as the primary/baseline representation used for the sensor-, descriptor-, and prayer-component attribution (ablation) studies and for the learned-representation comparison.

%whereas Eq.~\eqref{eq:qibla_six} preserves directional information after physical canonicalization. The two representations test different invariance assumptions rather than defining an ``old'' and ``new'' preprocessing route.

\subsection{Qibla-Referenced IMU Representation}

The Qibla-referenced IMU representation is obtained through a number of steps as follows. 

%\textit{Cross-Platform Sensor Convention Harmonization:}
%Let $\mathbf a_D(t)$, $\boldsymbol\omega_D(t)$, and $\mathbf g_D(t)$ denote acceleration, angular velocity, and gravity expressed in the device-fixed frame $D$. Sensor Logger documents platform-specific acceleration/gravity and orientation conventions \cite{sensorlogger_crossplatform,sensorlogger_orientation}.

\textit{Orientation Reconstruction and Native-Rate Synchronization:}
The Sensor Logger app reports yaw, pitch, and roll under an intrinsic $Z$--$X'$--$Y''$
Tait--Bryan convention \cite{sensorlogger_orientation}. We construct a proper
device-to-platform-world rotation $\mathbf{R}_{P\leftarrow D}(t)$ from the recorded Euler
angles using the platform-specific direction convention. Native
non-standardized iOS uses $(\psi,\theta,\phi)$ directly, whereas Android and
standardized iOS use $(-\psi,-\theta,\phi)$ before constructing the
$Z$--$X'$--$Y''$ rotation.

Euler angles are not interpolated directly. Each orientation sample is converted
to a unit quaternion $\mathbf{q}=[q_w,q_x,q_y,q_z]^{\top}$ and adjacent quaternion signs are made antipodally
continuous. The resulting rotation trajectory is evaluated independently at the
acceleration, gravity, and gyroscope timestamps using spherical linear interpolation (SLERP)
\cite{shoemake1985slerp}; when a sensor timestamp coincides with a recorded
orientation timestamp, the measured orientation is recovered directly.
Interpolation is not bridged across orientation gaps exceeding 0.25~s. 

\textit{Device-frame to World-frame rotation:}
A vector $\tilde{\mathbf x}_D\in\{\tilde{\mathbf a}_D,\tilde{\boldsymbol\omega}_D\}$
in the harmonized device frame is then expressed in the platform world frame as
\begin{equation}
\mathbf x_P(t)=\mathbf{R}_{P\leftarrow D}(t)\tilde{\mathbf x}_D(t).
\label{eq:device_to_platform}
\end{equation}
with the rotation matrix given as
\begin{equation}
\setlength{\arraycolsep}{2pt}
\mathbf{R}_{P\leftarrow D}=
\begin{bmatrix}
1-2(q_y^2+q_z^2) & 2(q_xq_y-q_wq_z) & 2(q_xq_z+q_wq_y)\\
2(q_xq_y+q_wq_z) & 1-2(q_x^2+q_z^2) & 2(q_yq_z-q_wq_x)\\
2(q_xq_z-q_wq_y) & 2(q_yq_z+q_wq_x) & 1-2(q_x^2+q_y^2)
\end{bmatrix}.
\end{equation}

\textit{True-North Harmonization:}
The platform world frame is north--west--up (NWU), but its horizontal reference
is platform dependent: the iOS orientation stream is nominally referenced to
true north, whereas Android is referenced to magnetic north
\cite{sensorlogger_orientation}. 
A globally consistent physical frame requires distinguishing true-north and magnetic-north orientation references.
For Android recordings, geomagnetic declination $\delta$ can be obtained from the World Magnetic
Model (WMM), the standard navigation/heading geomagnetic model \cite{chulliat2025wmm}. 
For the collection period, the geomagnetic declination $\delta$ is approximately $3.68^{\circ}$, for the KAUST site. The platform-world vector ${\mathbf x}_P\in\{{\mathbf a}_P,{\boldsymbol\omega}_P\}$ is mapped to a common true-north NWU frame $T$ as
\begin{equation}
\mathbf x_T(t)=\mathbf{R}_{T\leftarrow P}(\delta)\mathbf x_P(t),
\label{eq:magnetic_true}
\end{equation}
with 
\begin{equation}
\mathbf{R}_{T\leftarrow P}(\delta)=
\begin{bmatrix}
\cos\delta & -\sin\delta & 0\\
\sin\delta & \cos\delta & 0\\
0 & 0 & 1
\end{bmatrix}
\label{eq:R_declination}
\end{equation}
for Android, and
$\mathbf{R}_{T\leftarrow P}=\mathbf{I}_3$ for iOS. 

Note that WMM is used only to remove the magnetic-versus-true-north reference difference; it does not reconstruct or
calibrate the handset magnetometer.

\textit{Qibla-Referenced Directional Representation:}
The IMU prayer data collection was done at KAUST Grand Mosque whose coordinates are as follows:
$(22.3120556^{\circ}\mathrm{N},39.1003333^{\circ}\mathrm{E})$. The coordinates of Qibla/Kaa'ba
are as follows: $(21.4224779^{\circ}\mathrm{N},39.8251832^{\circ}\mathrm{E})$.
Using the initial great-circle bearing, the Qibla direction from the KAUST mosque is $\beta = 142.767884^{\circ}$ clockwise from true north. In the true-NWU frame, define the orthonormal basis
\begin{equation}
\mathbf q=
\begin{bmatrix}\cos\beta\\-\sin\beta\\0\end{bmatrix},\quad
\mathbf l=
\begin{bmatrix}\sin\beta\\\cos\beta\\0\end{bmatrix},\quad
\mathbf v=
\begin{bmatrix}0\\0\\1\end{bmatrix},
\label{eq:qlv_basis}
\end{equation}
where $\mathbf q$ points toward the Qibla, $\mathbf l$ points horizontally left
when facing the Qibla, and $\mathbf v$ points vertically upward. The true-NWU to
Qibla-left-up rotation is
\begin{equation}
\mathbf{R}_{Q\leftarrow T}=
\begin{bmatrix}
\mathbf q^{\top}\\
\mathbf l^{\top}\\
\mathbf v^{\top}
\end{bmatrix}.
\label{eq:true_to_qibla}
\end{equation}
The complete canonicalization is therefore
\begin{equation}
\mathbf x_{QLV}(t)=
\mathbf{R}_{Q\leftarrow T}(\mathbf{R}_{T\leftarrow P}(\mathbf{R}_{P\leftarrow D}(t)
\tilde{\mathbf x}_D(t))),
\label{eq:full_qibla_transform}
\end{equation}
where $\mathbf x_{QLV}=[a_Q,a_L,a_V,\omega_Q,\omega_L,\omega_V]^T$, which represents the six Qibla-referenced directional channels due to acceleration and angular velocity. 

\begin{figure*}[!t]
\centering
\includegraphics[width=0.98\textwidth]{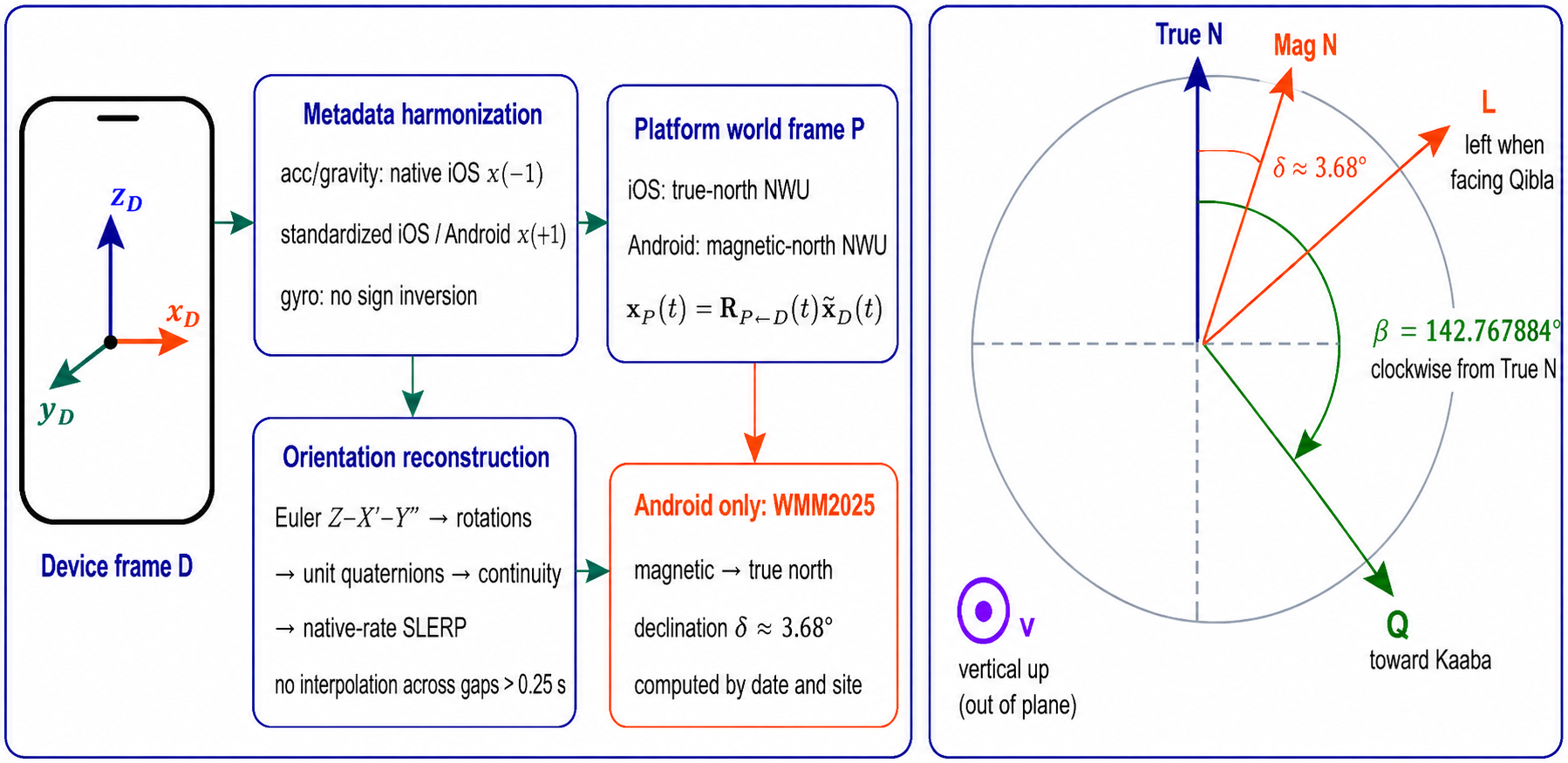}
\caption{Metadata-aware coordinate canonicalization. (a) Platform metadata
harmonize vector and orientation conventions; Euler-derived rotations are
synchronized to the native IMU timelines by quaternion SLERP. Android's
magnetic-north world reference is corrected to true north using WMM2025.
(b) True-NWU vectors are rotated into a Qibla-left-up basis, where $Q$ points
toward the Kaa'ba, $L$ is horizontal left when facing Qibla, and $V$ is vertical
up.}
\label{fig:qibla_canonicalization}
\end{figure*}

Fig. \ref{fig:qibla_canonicalization} provides a compact visual summary of Qibla-referenced directional IMU representation method used in this work.

%Proper-rotation, quaternion-norm, vector-norm, and transformed-gravity checks are used as physical implementation audits. 
%no participant-specific orientation, sign, or axis choice is selected using biometric scores.

%%%

\section{Biometric Evaluation Protocols}
\label{sec:protocols}

%The evaluation separates identification from verification and separates unit-level distinctiveness from session-level generalization. These distinctions are essential because multiple two-\rakaah{} units originate from the same continuous recording and therefore share acquisition conditions.

%Behavioral biometrics are highly context-dependent; variations arise from device handling, environment, physical state, and enrollment design \cite{alzubaidi2016authentication,abuhamad2020sensor,mahfouz2018survey}, and performance shifts with altered conditions \cite{siirtola2019context}. Also, standard random data splits are flawed when multiple samples come from one recording. 

%We therefore distinguish two evaluation protocols: the \emph{unit-level evaluation} (assessing participant structure within complete two-rakaah units) and the \emph{complete-session holdout evaluation} (testing generalization when query data is entirely absent during training). This separation (which also distinguishes identification from verification) is essential because units sharing acquisition conditions must be evaluated for true session-level generalization.

\subsection{Feature Extraction}

{Feature extraction} is an important pre-processing step for biometric analysis.
Each complete unit is divided into 5-s windows with 50\% overlap. Twenty
descriptors are computed per channel, spanning central tendency, variability,
range, distribution shape, first-difference dynamics, and spectral structure
(see Appendix A). The \emph{rotation-invariant
representation} (of section IV-A) therefore yields 100 features per window (corresponding to five channels), while the \emph{Qibla-referenced representation} (of section IV-B) yields 120 features per window (corresponding to six channels). For a unit with $M_u$ valid windows, window feature vectors $\mathbf h_{u,m}$ are averaged to form a single unit-level feature vector $\mathbf u=\frac{1}{M_u}\sum_{m=1}^{M_u}\mathbf h_{u,m}$.

%The broad signal-, descriptor-, and prayer-component ablations use the rotation-invariant representation. Their purpose is to attribute biometric information to sensing modalities and movement components while holding coordinate treatment fixed. The Qibla-referenced representation is used separately for the matched representation comparison and, after that comparison is established, for a physically interpretable directional ablation of the $Q$, $L$, and $V$ axes. 

\subsection{Terminology}
\label{sec:terminology_protocols}

We use the following terms in their standard biometric-evaluation sense
throughout the remainder of the paper.
\begin{itemize}[leftmargin=*]
\item \textit{Enrollment} is the set of behavioral samples used to represent a
participant's identity (a gallery entry). A \textit{query} is the sample being matched against enrollment.
\item \textit{Identification} assigns a query to one participant identity
(``who is this?''). \textit{Verification} tests a proposed identity claim for
one query--enrollment pair (``is this the claimed person?''). \textit{Pairwise
verification} specifically denotes verification performed on explicit
query--enrollment \emph{pairs} scored either by a fixed similarity (fixed-score
verification) or by a trained binary classifier (ML model-based pairwise
verification).
\item \textit{Closed-set} identification assumes every query belongs to one of
the enrolled identities and asks only which one. \textit{Open-set}
identification additionally allows a query from a non-enrolled identity and
must support rejection. All identification results in this paper are
closed-set; open-set rejection is not evaluated.
\item \textit{Unit-level analysis} treats every complete two-\rakaah{} unit,
regardless of which recording it came from, as an independent sample, so
units from the same recording may appear on both sides of a train/query split.
\textit{Cross-session} (complete-session) \textit{analysis} instead resamples
whole recording sessions, guaranteeing that every unit from the query
recording is absent from enrollment and training.
The two are complementary: unit-level analysis measures raw participant
distinctiveness, while cross-session analysis measures whether that
distinctiveness survives when the query is drawn from an entirely unseen
recording. 
%(Sec.~\ref{sec:unit_vs_session} below elaborates the two concrete protocols).
\end{itemize}

%Behavioral measurements are inherently context dependent. Reviews of smartphone authentication identify device handling, environment, physical state, and enrollment design as recurring sources of within-person variation \cite{alzubaidi2016authentication,mahfouz2018survey,abuhamad2020survey}. Context-specific work likewise shows that authentication performance can change when the same user is observed under altered conditions \cite{siirtola2019context}. This issue is especially important when multiple samples originate from one long recording: in such situations, random or unit-level splits can expose training and testing to a common acquisition state.

%We therefore distinguish two questions rather than treating them as interchangeable experiments. The unit-level protocol asks whether complete two-\rakaah{} units contain learnable participant structure and supports broad representation analysis. Complete-session holdout asks whether that structure persists when the query recording is entirely absent from enrollment and model fitting. This separation is essential for interpreting motion biometrics, where a representation can be highly discriminative within the collected data while remaining sensitive to session-dependent placement, orientation, or sensor conditions.

%The evaluation separates identification from verification and separates unit-level distinctiveness from session-level generalization. These distinctions are essential because multiple two-\rakaah{} units originate from the same continuous recording and therefore share acquisition conditions.

\subsection{Unit-Level Protocol}
In this evaluation protocol, every complete unit is treated as one behavioral sample. This evaluation branch uses the
full eligible cohort and is the principal setting for representation comparison,
ablations, and self-supervised analysis. The following evaluations are carried out at the unit-level. 

%\begin{equation}
%\mathbf u=\frac{1}{M_u}\sum_{m=1}^{M_u}\mathbf h_{u,m}.
%\label{eq:unitprototype}
%\end{equation}

%Likewise, the broad learned-model comparison holds the invariant input fixed, whereas a targeted SimCLR extension tests whether physically canonical directions alter self-supervised representation quality without repeating every neural objective under both inputs.

\subsubsection{Nearest-Neighbor Identification}
Each eligible unit is queried once and excluded from its own gallery. The geometric/distance metrics, namely, Pearson
correlation, cosine similarity, and Euclidean distance are evaluated.
%with negative Euclidean distance used so that larger scores indicate closer matches.
%Top-$k$ identification is correct when at least one unit from the true participant appears among the $k$ highest-ranked gallery samples.

\subsubsection{ML Model-based Identification}
Logistic Regression, Linear SVM, and Random Forest learn participant class
boundaries under leave-one-unit-out evaluation. The training is done on non-query data within the corresponding fold. This task differs
from nearest-neighbor retrieval because identity is assigned by a learned
multiclass decision function rather than by the closest sample.

\subsubsection{Fixed-Score Pairwise Verification}
Pairs of units are labeled genuine under hypothesis
$H_G$ when they belong to the same participant and impostor under hypothesis $H_I$ otherwise.
Pearson, cosine, and Euclidean scores are evaluated without learning a matching
function. 
%The receiver operating characteristic-area under the curve (ROC-AUC) metric measures threshold-independent ranking, while equal-error rate (EER), the operating point at which false-accept and false-reject rates are equal, is also recorded.

\subsubsection{ML Model-based Pairwise Verification}
For unit vectors $\mathbf u_a,\mathbf u_b\in\mathbb R^D$, the learned pair
representation is
\begin{equation}
\begin{aligned}
\phi(\mathbf u_a,\mathbf u_b)=\big[&|\mathbf u_a-\mathbf u_b|,
\mathbf u_a\odot\mathbf u_b,\\
&\rho(\mathbf u_a,\mathbf u_b),\;c(\mathbf u_a,\mathbf u_b),\;
 d_E(\mathbf u_a,\mathbf u_b)\big].
\end{aligned}
\label{eq:pairwise}
\end{equation}
where $\odot$ is the element-wise product and the final three terms are Pearson
correlation, cosine similarity, and Euclidean distance. Random Forest is the
principal learned verifier; reported point estimates are averaged over random
seeds 42, 123, and 2026.

\subsubsection{Participant Templates}
This method follows a leave-one-unit-out evaluation approach whereby the query unit is removed while the model
is fitted on all the remaining/enrollment units of that participant which are averaged to realize a participant template. This way, the query is compared against participant templates (with one template per participant).
This reduces unit-specific variation and corresponds more closely to biometric
enrollment than single-unit gallery matching.

\textit{Remark:}
The unit-level protocol is intentionally broad. Because units from a common
recording may occur on both sides of a fold, it demonstrates distinctiveness and
representation learnability but is not interpreted by itself as evidence of
cross-session biometric persistence.

\subsection{Complete-Session Holdout Protocol}
This is a more strict protocol that defines a behavioral sample as one entire nightly recording session. In each fold, every unit from one recording is
held out simultaneously. No query-session information is used to train ML classifiers, construct galleries or
participant templates, or generate learned training pairs. A recording is query
eligible only when its participant retains at least one independent enrollment
recording.

Nearest-neighbor queries are matched only against units from other recordings.
Participant templates are formed from non-query sessions. ML classifiers
are fitted to all non-query recordings and applied to every unit in the held-out
session. For ML model-based verification, genuine training pairs must span two sessions
from the same participant, impostor pairs join different participants, and any
pair containing a query unit is excluded from training. Up to 20,000 training
pairs per fold are used by the efficient pairwise Random Forest implementation,
with 100 trees and the same three seeds. The resulting complete-session
population contains 32 query-eligible participants, 132 held-out recordings,
and 527 query units; single-session participants can remain as competing
identities but cannot serve as query identities.

\subsection{Learned-Representation Evaluation}
The learned temporal representation evaluation branch consists of supervised and self-supervised learning models that utilize the rotation-invariant five-channel IMU representation. 
%The purpose is to isolate the effect of temporal representation learning from the effect of coordinate canonicalization. 
The Window CNN, full-unit CNN, SimCLR, BYOL, and DINO are first compared under the unit-level
protocol; the Window CNN and SimCLR are then evaluated under complete-session
holdout using identical session partitions and enrollment definitions.
One-dimensional CNNs are studied because they could capture local inertial dynamics, while
self-supervised learning tests whether useful participant structure can be
formed before identity labels are introduced. 

%Here, learned models provide complementary evidence to the controlled handcrafted analyses rather than replacing them.

\section{Main Biometric Results}
\label{sec:results}

\subsection{Performance Metrics}
\label{sec:metrics_defs}

Before presenting results we define every metric used below.
Identification is reported by \textit{Top-$k$ accuracy}, which is the fraction of queries for which at least one
true-participant sample appears among the $k$ highest-ranked gallery entries; we report Top-1 unless stated
otherwise. 
\textit{ML model-based accuracy} is the fraction of queries correctly
assigned by a trained multiclass classifier. Verification is reported by \textit{ROC AUC} (area under the
receiver operating characteristic curve) that summarizes verification ranking
quality independent of any operating threshold, with 1.0 denoting perfect
genuine/impostor separation and 0.5 denoting chance. \textit{EER} (equal error
rate) is the verification operating point at which the false-accept rate
equals the false-reject rate; lower EER indicates better separation. Where
used, \textit{macro F1} averages the per-class F1 score unweighted by class
size. 
When score separation is diagnostically useful, the decidability index summarizes genuine/impostor score separation in standardized units: 
\begin{equation}
 d' = \frac{|\mu_g-\mu_i|}{\sqrt{(\sigma_g^2+\sigma_i^2)/2}},
\end{equation}
where $(\mu_g,\sigma_g)$ and $(\mu_i,\sigma_i)$ denote genuine and impostor score statistics. 

Where confidence intervals are shown, 95\% intervals are obtained by participant-clustered bootstrap resampling. This preserves dependence among repeated observations from the same person. All point estimates for the learned pairwise Random Forest are averaged over three random seeds.

\subsection{Matched Representation Comparison}
Table~\ref{tab:representation_comparison} summarizes the main handcrafted
results for the unit-level analysis and complete-session holdout analysis, under two IMU representations. 

\begin{table*}[!t]
\centering
\caption{Main behavioral biometrics results: Matched handcrafted identification and verification results for the
rotation-invariant magnitude/gravity-relative and Qibla-referenced directional
IMU representations. Each row reports the best classifier or fixed (similarity-based) rule for
that task. Learned pairwise Random Forest point estimates are averaged across
three random seeds. LR: Logistic regression, SVM: support vector machine, RF: random forest.}
\label{tab:representation_comparison}
\footnotesize
\setlength{\tabcolsep}{5pt}
\renewcommand{\arraystretch}{1.12}
\begin{tabularx}{\textwidth}{>{\raggedright\arraybackslash}p{2.4cm}>{\raggedright\arraybackslash}p{3.0cm}>{\raggedright\arraybackslash}X>{\raggedright\arraybackslash}X}
\toprule
Protocol & Evaluation task & Rotation-invariant IMU representation & Qibla-referenced IMU representation\\
\midrule
\multirow{5}{*}{Unit-level}
& Nearest-neighbor identification & Pearson Top-1: 86.1\% & Pearson/Cosine Top-1: 87.16\%\\
& ML model-based identification & Linear SVM accuracy: 95.9\% & Linear SVM accuracy: 96.57\%\\
& Fixed-score verification & Pearson AUC/EER: 0.827/23.9\% & Pearson AUC/EER: 0.8418/24.33\%\\
& ML model-based pairwise verification & RF AUC/EER: 0.9932/4.07\% & RF AUC/EER: 0.9923/3.96\%\\
& Participant-template matching & Pearson AUC/EER/Top-1: 0.947/11.18\%/66.7\% & Pearson AUC/EER/Top-1: 0.9506/11.21\%/69.85\%\\
\midrule
\multirow{5}{*}{Complete-session}
& Nearest-neighbor identification & Euclidean Top-1: 72.9\% & Cosine Top-1: 72.52\%\\
& ML model-based identification & Random Forest accuracy: 77.8\% & Logistic Regression accuracy: 80.60\%\\
& Fixed-score verification & Pearson AUC/EER: 0.814/24.82\% & Pearson AUC/EER: 0.8269/25.83\%\\
& ML model-based pairwise verification & RF AUC/EER: 0.9700/7.62\% & RF AUC/EER: 0.9618/7.81\%\\
& Participant-template matching & Pearson AUC/EER/Top-1: 0.906/15.94\%/55.4\% & Pearson AUC/EER/Top-1: 0.9081/14.67\%/54.97\%\\
\bottomrule
\end{tabularx}
\end{table*}

The unit-level protocol establishes strong participant distinctiveness for IMU both
representations. The Qibla-referenced representation slightly increases
nearest-neighbor and ML model-based identification, reaching an accuracy of 87.16\% and 96.57\%,
respectively. Fixed Pearson AUC also increases from 0.827 to 0.8418, while the
ML model-based pairwise verifier is effectively saturated for both representations:
0.9932/4.07\% AUC/EER for the invariant representation and 0.9923/3.96\% for
the directional representation. Template aggregation is similarly strong,
with Qibla-referenced AUC 0.9506 and Top-1 69.85\%.

Complete-session holdout provides the more demanding biometric test. The
Qibla-referenced representation improves the best ML model-based accuracy
from 77.8\% to 80.60\% and fixed Pearson AUC from 0.814 to 0.8269. In contrast,
nearest-neighbor Top-1 and ML model-based pairwise verification are similar or slightly
stronger for the invariant route; the latter obtains AUC/EER 0.9700/7.62\%
versus 0.9618/7.81\%. Participant-template verification is comparable, with
AUC 0.906 and 0.9081, respectively. Thus, physical directional canonicalization
is informative but not uniformly superior. The key result is that substantial
cross-session identity structure survives under both an analytically invariant
representation and an explicitly canonical physical frame.

%The rotation-invariant representation is the analysis backbone used later for attribution and learned-representation experiments; the Qibla-referenced route is the directional canonicalization experiment.

%\begin{figure}[!t]
%\centering
%\includegraphics[width=\columnwidth]{figures/figure4_fourcurve_original_and_strict_roc.png}
%\caption{ROC curves for fixed Pearson similarity and learned pairwise Random Forest verification under the unit-level and complete-session protocols for the rotation-invariant handcrafted representation.}
%\label{fig:handcrafted_roc_comparison}
%\end{figure}

\section{Where Is the Identity Information?}
\label{sec:identity_information}

%As mentioned in previous section, the three attribution experiments in this section use the rotation-invariant magnitude/gravity-relative representation. 
%This is intentional: holding coordinate treatment fixed allows differences across signal families, descriptor families, and prayer components to be interpreted as differences in information content rather than as consequences of changing the global horizontal reference. 
%A final, complementary analysis then exploits the physical meaning of the Qibla-referenced frame to ask whether identity information is concentrated along the Qibla direction, the lateral direction, the vertical direction, or particular sensor families.

%The representation comparison also motivates the division of labor used in the remainder of the paper. 
The rotation-invariant representation is a controlled backbone
for asking \emph{where} biometric information resides while doing sensor-, descriptor-, or
prayer-component ablation analysis. The Qibla-referenced route answers the separate
question of whether preserving canonical direction adds biometric value.

\subsection{IMU Sensor-Family Ablation}
\begin{table*}[t]
\centering
\caption{IMU Signal-family ablation of the rotation-invariant magnitude/gravity-relative IMU representation under the unit-level and complete-session protocols. For each signal representation, the
reported ML model-based accuracy is the highest result among logistic regression, linear SVM, and random forest, with the selected classifier
shown beneath the value. The complete-session fixed-score AUC is the highest
among Pearson, cosine, and Euclidean scoring, with the selected rule shown
beneath the value. 
%Best ML-model based accuracy: best identification accuracy among LR, SVM, and RF. 
}
\label{tab:sensor_ablation}
\resizebox{\textwidth}{!}{%
\begin{tabular}{lcccccccc}
\toprule
&
&
\multicolumn{3}{c}{\textbf{Unit-level protocol}}
&
\multicolumn{4}{c}{\textbf{Complete-session holdout protocol}}
\\
\cmidrule(lr){3-5}
\cmidrule(lr){6-9}

\textbf{Signal representation}
&
\textbf{\# features}
&
\shortstack{\textbf{Best ML model-based}\\\textbf{accuracy (\%)}}
&
\shortstack{\textbf{Pairwise RF}\\\textbf{AUC}}
&
\shortstack{\textbf{Pairwise RF}\\\textbf{EER (\%)}}
&
\shortstack{\textbf{Best  ML model-based}\\\textbf{accuracy (\%)}}
&
\shortstack{\textbf{Best fixed-score}\\\textbf{AUC}}
&
\shortstack{\textbf{Pairwise RF}\\\textbf{AUC}}
&
\shortstack{\textbf{Pairwise RF}\\\textbf{EER (\%)}}
\\
\midrule

\multicolumn{9}{l}{\textit{Reference representation}}\\[1pt]

All five derived signals
&
100
&
\shortstack{\textbf{95.9}\\{\scriptsize SVM}}
&
\textbf{0.9919}
&
4.89
&
\shortstack{\textbf{78.0}\\{\scriptsize RF}}
&
\shortstack{0.8140\\{\scriptsize Pearson}}
&
\textbf{0.9703}
&
\textbf{7.63}
\\

\addlinespace[3pt]
\multicolumn{9}{l}{\textit{Two-signal-family representations}}\\[1pt]

Acceleration + gravity-derived
&
80
&
\shortstack{93.8\\{\scriptsize SVM}}
&
0.9899
&
\textbf{4.85}
&
\shortstack{77.8\\{\scriptsize RF}}
&
\shortstack{0.8147\\{\scriptsize Pearson}}
&
0.9691
&
7.74
\\

Acceleration + gyroscope magnitudes
&
40
&
\shortstack{87.3\\{\scriptsize SVM}}
&
0.9855
&
5.80
&
\shortstack{77.8\\{\scriptsize LR}}
&
\shortstack{0.7866\\{\scriptsize Pearson}}
&
0.9448
&
11.25
\\

Gyroscope + gravity magnitudes
&
40
&
\shortstack{76.5\\{\scriptsize RF}}
&
0.9834
&
6.45
&
\shortstack{68.1\\{\scriptsize RF}}
&
\shortstack{0.7682\\{\scriptsize Pearson}}
&
0.9604
&
9.21
\\

\addlinespace[3pt]
\multicolumn{9}{l}{\textit{Single-signal representations}}\\[1pt]

Acceleration magnitude only
&
20
&
\shortstack{73.0\\{\scriptsize SVM}}
&
0.9785
&
7.08
&
\shortstack{60.3\\{\scriptsize SVM}}
&
\shortstack{0.7626\\{\scriptsize Euclidean}}
&
0.9314
&
13.02
\\

Gravity magnitude only
&
20
&
\shortstack{62.6\\{\scriptsize RF}}
&
0.9758
&
8.05
&
\shortstack{60.7\\{\scriptsize RF}}
&
\shortstack{\textbf{0.8358}\\{\scriptsize Pearson}}
&
0.9566
&
10.41
\\

Gyroscope magnitude only
&
20
&
\shortstack{57.7\\{\scriptsize SVM}}
&
0.9344
&
13.76
&
\shortstack{51.4\\{\scriptsize RF}}
&
\shortstack{0.7087\\{\scriptsize Pearson}}
&
0.8921
&
17.96
\\

\bottomrule
\end{tabular}%
}
\end{table*}

Table \ref{tab:sensor_ablation} provides results for IMU signal-family ablation. 
Acceleration plus gravity-derived channels remain close to the full
representation for ML model-based verification, whereas gyroscope magnitude is weak in
isolation. Gravity magnitude is noteworthy: despite being nominally close to
local gravitational acceleration under quasi-static conditions, it alone
produces nontrivial discrimination, including complete-session fixed-score AUC
0.8358. This result is not interpreted as evidence that a physical constant is
a human biometric. The gravity stream contains motion-dependent sensor-fusion
behavior, while persistent device bias, scale factor, noise, and platform
processing may also contribute. The issue is revisited in
Sec.~\ref{sec:device_confound}.

\subsection{Descriptor-Family Ablation}
\begin{table*}[t]
\centering
\caption{Descriptor-family ablation using the rotation-invariant magnitude/gravity-relative IMU representation under the unit-level and complete-session protocols.}
\label{tab:feature_group_ablation}
\resizebox{\textwidth}{!}{%
\begin{tabular}{lcccccccc}
\toprule
&
&
\multicolumn{3}{c}{\textbf{Unit-level protocol}}
&
\multicolumn{4}{c}{\textbf{Complete-session holdout protocol}}
\\
\cmidrule(lr){3-5}
\cmidrule(lr){6-9}

\textbf{Feature representation}
&
\textbf{\# features}
&
\shortstack{\textbf{Best ML model-based}\\\textbf{accuracy (\%)}}
&
\shortstack{\textbf{Pairwise RF}\\\textbf{AUC}}
&
\shortstack{\textbf{Pairwise RF}\\\textbf{EER (\%)}}
&
\shortstack{\textbf{Best model-based}\\\textbf{accuracy (\%)}}
&
\shortstack{\textbf{Best fixed-score}\\\textbf{AUC}}
&
\shortstack{\textbf{Pairwise RF}\\\textbf{AUC}}
&
\shortstack{\textbf{Pairwise RF}\\\textbf{EER (\%)}}
\\
\midrule

\multicolumn{9}{l}{\textit{Reference representation}}\\[1pt]

All targeted features
&
100
&
\shortstack{\textbf{95.9}\\{\scriptsize SVM}}
&
\textbf{0.9919}
&
4.89
&
\shortstack{78.0\\{\scriptsize RF}}
&
\shortstack{0.8140\\{\scriptsize Pearson}}
&
\textbf{0.9703}
&
\textbf{7.63}
\\

\addlinespace[3pt]
\multicolumn{9}{l}{\textit{Descriptor-family subsets}}\\[1pt]

Time-domain features
&
90
&
\shortstack{92.0\\{\scriptsize SVM}}
&
0.9874
&
5.35
&
\shortstack{74.6\\{\scriptsize LR}}
&
\shortstack{0.7942\\{\scriptsize Pearson}}
&
0.9646
&
8.65
\\

Central tendency and variability
&
75
&
\shortstack{90.5\\{\scriptsize SVM}}
&
0.9851
&
5.55
&
\shortstack{73.4\\{\scriptsize SVM}}
&
\shortstack{0.7658\\{\scriptsize Pearson}}
&
0.9626
&
8.87
\\

Motion-dynamics features
&
20
&
\shortstack{87.7\\{\scriptsize SVM}}
&
0.9900
&
\textbf{4.86}
&
\shortstack{76.3\\{\scriptsize RF}}
&
\shortstack{0.8458\\{\scriptsize Euclidean}}
&
0.9674
&
8.37
\\

Frequency-domain features
&
10
&
\shortstack{80.1\\{\scriptsize RF}}
&
0.9878
&
6.15
&
\shortstack{73.1\\{\scriptsize RF}}
&
\shortstack{\textbf{0.8958}\\{\scriptsize Pearson}}
&
0.9594
&
9.34
\\

Magnitude features
&
60
&
\shortstack{91.0\\{\scriptsize SVM}}
&
0.9911
&
5.11
&
\shortstack{\textbf{78.8}\\{\scriptsize LR}}
&
\shortstack{0.7986\\{\scriptsize Pearson}}
&
0.9673
&
7.92
\\

\bottomrule
\end{tabular}%
}
\end{table*}

Table \ref{tab:feature_group_ablation} provides results for feature-family ablation. 
The complete feature set remains strongest overall for ML model-based pairwise
verification, but compact feature families retain substantial performance.
Motion-dynamics descriptors provide strong performance at low dimensionality,
while the 10 frequency-domain features produce the highest complete-session
fixed-score AUC (0.8958). The latter indicates that cross-session identity
structure is expressed not only in average movement magnitude but also in rhythm
and temporal regularity.

\subsection{Prayer-Component Ablation}
\begin{table*}[t]
\centering
\caption{Selected prayer-component ablation results using the rotation-invariant magnitude/gravity-relative IMU representation under the unit-level and complete-session protocols. Note: A \emph{Prostration phase} consists of two prostration (Sujud) events and the sitting (Julus) event between them.}
\label{tab:posture_group_ablation}
\resizebox{\textwidth}{!}{%
\begin{tabular}{lcccccccc}
\toprule
&
&
\multicolumn{3}{c}{\textbf{Unit-level protocol}}
&
\multicolumn{4}{c}{\textbf{Complete-session holdout protocol}}
\\
\cmidrule(lr){3-5}
\cmidrule(lr){6-9}

\textbf{Posture representation}
&
\shortstack{\textbf{Median retained}\\\textbf{duration (s)}}
&
\shortstack{\textbf{Best ML model-based}\\\textbf{accuracy (\%)}}
&
\shortstack{\textbf{Pairwise RF}\\\textbf{AUC}}
&
\shortstack{\textbf{Pairwise RF}\\\textbf{EER (\%)}}
&
\shortstack{\textbf{Best ML model-based}\\\textbf{accuracy (\%)}}
&
\shortstack{\textbf{Best fixed-score}\\\textbf{AUC}}
&
\shortstack{\textbf{Pairwise RF}\\\textbf{AUC}}
&
\shortstack{\textbf{Pairwise RF}\\\textbf{EER (\%)}}
\\
\midrule

\multicolumn{9}{l}{\textit{Reference representation}}\\[1pt]

All labeled posture intervals
&
\shortstack{447\\{\scriptsize (100\%)}}
&
\shortstack{89.7\\{\scriptsize SVM}}
&
\textbf{0.9913}
&
4.96
&
\shortstack{\textbf{78.2}\\{\scriptsize RF}}
&
\shortstack{0.8009\\{\scriptsize Euclidean}}
&
0.9688
&
\textbf{7.27}
\\

\addlinespace[3pt]
\multicolumn{9}{l}{\textit{Posture-related subsets}}\\[1pt]

Prostration phase-related
&
\shortstack{101\\{\scriptsize (22.6\%)}}
&
\shortstack{\textbf{93.8}\\{\scriptsize LR}}
&
0.9778
&
6.31
&
\shortstack{78.0\\{\scriptsize RF}}
&
\shortstack{0.8450\\{\scriptsize Pearson}}
&
0.9693
&
8.06
\\

Sujud only
&
\shortstack{56\\{\scriptsize (12.5\%)}}
&
\shortstack{92.3\\{\scriptsize SVM}}
&
0.9864
&
6.33
&
\shortstack{\textbf{78.2}\\{\scriptsize RF}}
&
\shortstack{0.8477\\{\scriptsize Pearson}}
&
\textbf{0.9697}
&
7.67
\\

Bowing-related
&
\shortstack{51\\{\scriptsize (11.4\%)}}
&
\shortstack{86.4\\{\scriptsize RF}}
&
0.9838
&
7.80
&
\shortstack{76.5\\{\scriptsize RF}}
&
\shortstack{\textbf{0.8543}\\{\scriptsize Pearson}}
&
0.9645
&
8.00
\\

Standing only
&
\shortstack{280\\{\scriptsize (62.6\%)}}
&
\shortstack{79.8\\{\scriptsize RF}}
&
0.9877
&
\textbf{4.95}
&
\shortstack{74.6\\{\scriptsize RF}}
&
\shortstack{0.7453\\{\scriptsize Pearson}}
&
0.9593
&
9.11
\\

\addlinespace[3pt]
\multicolumn{9}{l}{\textit{Transition subset}}\\[1pt]

Transitions
&
\shortstack{23\\{\scriptsize (5.1\%)}}
&
\shortstack{78.8\\{\scriptsize RF}}
&
0.9777
&
7.97
&
\shortstack{72.1\\{\scriptsize RF}}
&
\shortstack{0.8123\\{\scriptsize Pearson}}
&
0.9664
&
8.73
\\

\bottomrule
\end{tabular}%
}
\end{table*}

Table \ref{tab:posture_group_ablation} provides results for prayer-component ablation. 
Prayer-component informativeness is not explained by duration alone. The
prostration-phase subset\footnote{A \emph{Prostration phase} consists of two prostration (Sujud) events and the sitting (Julus) event between them.} occupies only 22.6\% of retained labeled duration yet
approaches the full representation for several cross-session metrics; Sujud-only
uses 12.5\% of the duration and matches the full labeled representation's
complete-session ML model-based accuracy. Bowing-related motion yields the strongest
fixed-score verification among the selected subsets, whereas standing is much
longer but weaker for nearest-neighbor and fixed-score matching. Participant
information is therefore concentrated in selected dynamic components of the
prescribed sequence rather than being proportional to observation time.

\subsection{Directional Attribution in the Qibla-Referenced Frame}
The Qibla-referenced IMU representation is a physically canonical representation that enables the directional attribution question (that is not available from the rotation-invariant representation). We therefore repeat the
handcrafted ablation analysis by retaining selected subsets of
$[a_Q,a_L,a_V,\omega_Q,\omega_L,\omega_V]$. The ablation includes all
acceleration axes, all angular-velocity axes, each physical direction with both
sensor types, and the horizontal $Q$--$L$ plane. The same 20 handcrafted
descriptors are retained per selected channel. 
%To avoid conflating this physical axis analysis with minor classifier-implementation differences across analysis notebooks, 

%for which the full-channel reference reproduces the main Qibla result closely.
%The full six-channel reference in the main experiment obtains unit-level Top-1 87.16\% and RF AUC/EER 0.9923/3.96\%, and complete-session Top-1 72.52\%, fixed Pearson AUC 0.8269, and RF AUC/EER 0.9618/7.81\%.

% INTERNAL: The subset values below currently come from the available Qibla
% directional-ablation run. Reconcile its cohort with the visible full-cohort
% drafting assumption before final submission.
\begin{table*}[t]
\centering
\caption{Directional ablation using the Qibla-referenced IMU representation, under the unit-level and complete-session protocols. Subset pairwise-RF values are means over seeds 42, 123, and 2026. NN: nearest neighbor. }
\label{tab:qibla_direction_ablation}
\footnotesize
\begin{adjustbox}{max width=\textwidth}
\begin{tabular}{lcccccc}
\toprule
\textbf{Qibla-frame subset}
& \textbf{\# feat.}
& \shortstack{\textbf{Unit NN}\\\textbf{Top-1 (\%)}}
& \shortstack{\textbf{Unit RF}\\\textbf{AUC / EER (\%)}}
& \shortstack{\textbf{Session NN}\\\textbf{Top-1 (\%)}}
& \shortstack{\textbf{Session fixed}\\\textbf{AUC}}
& \shortstack{\textbf{Session RF}\\\textbf{AUC / EER (\%)}}\\
\midrule
Acceleration, $Q+L+V$ & 60 & 80.30 & 0.9903 / 4.57 & 66.28 & 0.8007 & 0.9560 / 8.25\\
Angular velocity, $Q+L+V$ & 60 & 73.13 & 0.9748 / 8.13 & 61.66 & 0.8115 & 0.9463 / 10.67\\
$Q$ direction, $a_Q+\omega_Q$ & 40 & 64.63 & 0.9649 / 9.88 & 53.35 & 0.8013 & 0.9358 / 13.11\\
$L$ direction, $a_L+\omega_L$ & 40 & 70.60 & 0.9757 / 8.06 & 59.35 & 0.7903 & 0.9470 / 10.84\\
$V$ direction, $a_V+\omega_V$ & 40 & \textbf{81.49} & \textbf{0.9884 / 5.31} & \textbf{68.36} & 0.7761 & \textbf{0.9487 / 9.42}\\
Horizontal $Q$--$L$ plane & 80 & 78.81 & 0.9819 / 6.81 & 65.36 & \textbf{0.8218} & 0.9545 / 9.95\\
\bottomrule
\end{tabular}
\end{adjustbox}
\end{table*}

Table~\ref{tab:qibla_direction_ablation} summarizes nearest-neighbor
identification, fixed-score verification, and ML model-based pairwise verification results, under the unit-level and complete-session protocols.
The ablation yields three consistent observations. First, no single physical
axis explains the complete result: retaining all six channels remains strongest
overall, indicating complementary information across direction and sensor type.
Second, acceleration preserves substantially more ML model-based pairwise performance
than angular velocity alone; under complete-session holdout, acceleration-only
RF verification reaches AUC/EER 0.9560/8.25\%, compared with
0.9463/10.67\% for angular velocity. Third, vertical motion is the strongest
single-direction representation for nearest-neighbor identification and
pairwise verification, whereas the $Q$ direction is the weakest single axis.
%Thus, Qibla canonicalization is not useful merely because forward/backward motion toward the prayer direction becomes explicit. 
Participant information is
distributed across vertical transitions, lateral/postural control, and
rotational dynamics. The horizontal $Q$--$L$ plane is nevertheless competitive
for fixed-score verification (AUC 0.8218), showing that canonical horizontal
structure also carries identity information.

%\section{Learned Temporal Representations}
%\section{CNN and SSL-based results}
\section{Supervised and Self-Supervised Learning (SSL)-based Results}
\label{sec:learned_results}

%The broad learned-model comparison holds the rotation-invariant five-channel input fixed. This isolates the effect of temporal representation learning from the effect of coordinate canonicalization: 

This section reports the results due to CNN and SSL models that utilize the rotation-invariant IMU representation and aim to learn participant-discriminative temporal structure. This section also reports the results of a SimCLR based experiment that utilizes the Qibla-referenced IMU representation.

%whereas the representation comparison in Sec.~\ref{sec:main_results} separately tests the value of retaining canonical direction.

A supervised full-unit 1-D CNN provides the baseline. The Window
CNN instead maps each 5-s window $\mathbf x_{u,m}\in\mathbb R^{C\times L}$,
with $C=5$, $L=512$ to an embedding $f_\theta(\mathbf x_{u,m})$ and averages embeddings
within the unit. The SSL branch uses the same windows but pretrains
without identity labels. SimCLR, BYOL, and DINO receive independently augmented
views formed through temporal cropping, amplitude scaling, additive noise,
channel dropout, temporal masking, and shifting. Participant identities are
introduced only during downstream retrieval, verification, and template
construction.

\begin{table*}[t]
\centering
\caption{Learned temporal representations under the unit-level protocol.
The supervised CNNs and SSL models use the
rotation-invariant five-channel input. Bold values in the SSL comparison denote the best result among SimCLR, DINO, and BYOL.
\emph{Note: The final block is a SimCLR evaluation using the six Qibla-referenced channels}. Qibla-referenced SimCLR results are averaged over the two completed random seeds (42 and 123); no two-sample standard deviation is reported. 
}
\label{tab:learned_original}
\footnotesize

\begin{adjustbox}{max width=\textwidth}
\begin{tabular}{lllccc}
\toprule
\textbf{Input}
& \textbf{Method}
& \textbf{Evaluation}
& \textbf{Identification result}
& \textbf{AUC}
& \textbf{EER (\%)}\\
\midrule

\multicolumn{6}{l}{\textit{Supervised temporal baselines: rotation-invariant five-channel input}}\\[1pt]

Invariant
& Full-unit CNN
& Unit classification
& Accuracy: $66.67 \pm 5.69\%$
& --
& --\\

Invariant
& Window CNN
& Unit classification and verification
& \shortstack{Accuracy: 78.25\%\\Macro F1: 70.23\%}
& 0.9331
& 13.38\\

\midrule

\multicolumn{6}{l}{\textit{SSL model comparison: rotation-invariant five-channel input}}\\[1pt]

\multirow{2}{*}{Invariant}
& \multirow{2}{*}{SimCLR}
& Unit-to-unit
& Top-1: \textbf{91.75\%}
& \textbf{0.8955}
& \textbf{16.57}\\

&
&
Participant template
& Top-1: \textbf{75.52\%}
& \textbf{0.9699}
& \textbf{8.52}\\

\addlinespace[1pt]

\multirow{2}{*}{Invariant}
& \multirow{2}{*}{DINO}
& Unit-to-unit
& Top-1: 85.82\%
& 0.8573
& 19.72\\

&
&
Participant template
& Top-1: 60.18\%
& 0.9473
& 12.39\\

\addlinespace[1pt]

\multirow{2}{*}{Invariant}
& \multirow{2}{*}{BYOL}
& Unit-to-unit
& Top-1: 77.45\%
& 0.8526
& 20.90\\

&
&
Participant template
& Top-1: 49.48\%
& 0.9247
& 15.12\\

\midrule

\multicolumn{6}{l}{\textit{SimCLR evaluation on Qibla-referenced six-channel input}}\\[1pt]

\multirow{2}{*}{Qibla-referenced}
& \multirow{2}{*}{SimCLR}
& Unit-to-unit
& Top-1: 90.96\%
& 0.9021
& 15.33\\

&
&
Participant template
& Top-1: 82.89\%
& 0.9811
& 6.03\\

\bottomrule
\end{tabular}
\end{adjustbox}

\end{table*}

Table~\ref{tab:learned_original} reports the unit-level results.
The Window CNN substantially outperforms the full-unit CNN, supporting local
pattern learning followed by aggregation rather than forcing the complete
variable-duration unit into one resampled trajectory. Among the rotation-invariant-input
SSL encoders, SimCLR is strongest: participant-template AUC/EER
reaches 0.9699/8.52\%, while unit-to-unit Top-1 reaches 91.75\%.

The last block of Table~\ref{tab:learned_original} provides results for the Qibla-referenced SimCLR experiment which tests a narrower question:
whether physically canonical directional channels can improve the representation
formed by the strongest SSL model, i.e., SimCLR. Each 5-s Qibla-referenced window contains
$[a_Q,a_L,a_V,\omega_Q,\omega_L,\omega_V]$ and is resampled to 512 samples.
Temporal crop/resize, amplitude scaling, additive noise, channel dropout,
temporal masking, and small temporal shifts are retained, but random axis
rotations, axis permutations, and sign flips are excluded because they would
destroy the physical meaning of the $Q$, $L$, and $V$ axes. Two 80-epoch runs
produce similar unit-to-unit performance and a consistent template advantage:
template Top-1 rises to 82.36--83.41\%, AUC to 0.9805--0.9817, and EER falls to
5.81--6.24\%. Direct unit-to-unit Top-1 remains comparable to the rotation-invariant
SimCLR result. This pattern suggests that Qibla-referenced temporal embeddings
benefit from aggregation across enrollment units.

% INTERNAL: Current Qibla-SimCLR results use the available unit-level Qibla
% window cache and two completed seeds. No strict Qibla-SimCLR result is claimed.
%The Qibla-referenced SimCLR experiment is intentionally interpreted only within the unit-level protocol. Complete-session learned evaluation is reported for the invariant input, where the query session is excluded from learned-model training and enrollment.

\begin{table}[t]
\centering
\caption{Principal learned temporal representations using the rotation-invariant five-channel input under complete-session holdout. U2U: Unit-to-unit, Rep.: Representation, PT: Participant template, W-CNN: window-CNN. }
\label{tab:learned_complete_session}
\footnotesize
\begin{adjustbox}{max width=\textwidth}
\begin{tabular}{llccc}
\toprule
\textbf{Encoder}
& \textbf{Rep.}
& \textbf{Top-1 (\%)}
& \textbf{AUC}
& \textbf{EER (\%)}\\
\midrule
\multirow{2}{*}{W-CNN}
& U2U
& \shortstack{65.58\\{\scriptsize [53.17, 75.05]}}
& \shortstack{\textbf{0.8976}\\{\scriptsize [0.7965, 0.9331]}}
& \shortstack{\textbf{15.90}\\{\scriptsize [11.87, 25.56]}}\\
& PT
& \shortstack{53.38\\{\scriptsize [35.95, 67.50]}}
& \shortstack{0.8994\\{\scriptsize [0.8285, 0.9492]}}
& \shortstack{15.06\\{\scriptsize [9.76, 23.10]}}\\
\midrule
\multirow{2}{*}{SimCLR}
& U2U
& \shortstack{\textbf{72.03}\\{\scriptsize [60.89, 80.13]}}
& \shortstack{0.8709\\{\scriptsize [0.7910, 0.9081]}}
& \shortstack{19.66\\{\scriptsize [14.81, 28.48]}}\\
& PT
& \shortstack{\textbf{58.56}\\{\scriptsize [44.63, 68.95]}}
& \shortstack{\textbf{0.9091}\\{\scriptsize [0.8476, 0.9516]}}
& \shortstack{\textbf{14.51}\\{\scriptsize [9.90, 20.45]}}\\
\bottomrule
\end{tabular}
\end{adjustbox}
\end{table}

Table~\ref{tab:learned_complete_session} presents the complete-session holdout results, under rotation-invariant IMU representation. SimCLR achieves higher unit-to-unit Top-1 than
the Window CNN (72.03\% versus 65.58\%), whereas the Window CNN provides
stronger unit-level verification. Participant-template aggregation improves
verification for both encoders, consistent with averaging suppressing
unit-specific variation. These strict results show that learned cross-session
identity structure is not restricted to the handcrafted descriptor definitions.
%they are not used to infer strict-session performance for the auxiliary Qibla-SimCLR extension.

\section{Discussion}
\label{sec:discussion}

\subsection{Unit-Level Versus Cross-Session Analysis}
The two evaluation protocols establish complementary biometric properties.
Unit-level results show that repeated complete prayer units are strongly
participant-discriminative and support detailed representation analysis.
Complete-session holdout shows that a substantial portion of this structure
persists when enrollment and query originate from independent recordings.
Identification generally degrades more strongly than fixed-score verification,
whereas the ML model-based pairwise verifier remains robust, retaining
session-specific structure and persistent participant information.

\subsection{Rotation Invariance Versus Physical Canonicalization}
%The coordinate experiment should not be interpreted as a winner-take-all comparison. The magnitude/gravity-relative IMU representation removes arbitrary sensor-axis rotation analytically through norms and gravity-relative projection, whereas the Qibla-referenced route preserves directional structure after mapping every recording to the same true-north and prayer-relative basis. 

The matched results show that Qibla-referenced IMU representation can improve selected ML model-based and fixed-score
metrics, but it does not uniformly improve ML model-based pairwise verification. The
directional ablation adds an important qualification: the $Q$ (Qibla) axis itself is
not the dominant biometric direction. Vertical motion is the strongest single
axis, acceleration contributes more than angular velocity for ML model-based verification,
and the complete six-channel representation is best overall. The
Qibla-referenced SimCLR experiment similarly improves template verification without
materially changing direct unit-to-unit identification. Together, these results
indicate that participant information is recoverable under two materially different coordinate assumptions.
%physical canonicalization is valuable because it organizes complementary movement components into a common frame, not because one prayer-facing direction alone explains identity. making it less plausible that separability is solely a consequence of arbitrary phone orientation.

\subsection{Why ML Model-based Pairwise Verification Is Strong}
A scalar similarity imposes one global geometry on all feature dimensions. The
learned pair representation exposes feature-wise absolute differences and
products in addition to Pearson, cosine, and Euclidean comparisons. Its large
advantage over fixed similarity under both protocols suggests that genuine and
impostor structure is distributed across complementary feature interactions.
The small variation across three Random Forest seeds further indicates that the
aggregate learned-verification result is not a single-seed artifact.

\subsection{Device Heterogeneity is a Potential Confound}
\label{sec:device_confound}
The heterogeneous-phone collection introduces the most important interpretive
confound. Consumer smartphone inertial sensors exhibit device-dependent bias,
scale factor, noise, and stochastic error
\cite{kos2016crossplatform,kos2016biofeedback,capuano2023smartphone}. If a
participant is repeatedly measured with the same handset, stable hardware
characteristics can correlate with identity. Complete-session holdout removes
recording overlap but does not create a device-disjoint protocol; the present
study therefore cannot quantify exactly how much of the score is attributable
to human movement versus stable handset characteristics.

The gravity-magnitude ablation makes this issue visible rather than merely
hypothetical. Because $\|\mathbf g\|$ should remain close to local gravitational
acceleration in an ideal calibrated system, its nontrivial biometric performance
could partly reflect calibration or sensor-fusion signatures. Yet direct gravity
magnitude cannot explain the complete effect: strong discrimination remains
when it is absent, including acceleration+gyroscope-magnitude pairwise AUC
0.9855 and Qibla-referenced six-axis ML model-based accuracy 96.57\% with learned
pairwise AUC 0.9923. Other accelerometer/gyroscope hardware characteristics may
still contribute, however.

A definitive behavioral attribution requires a crossed participant--device
design: the same participant recorded with multiple handset models and multiple
participants recorded with the same calibrated reference handset. Such data
would permit device-disjoint evaluation and direct separation of hardware
persistence from behavioral persistence.

\subsection{Security Interpretation}
The biometric experiments primarily model a \emph{zero-effort impostor}: an
unauthorized user naturally performs the same prescribed prayer sequence rather
than deliberately imitating a target. Learned genuine--impostor verification
therefore measures identity separability under ordinary behavioral variation,
not resistance to targeted imitation, replay, sensor injection, or adversarial
spoofing. Likewise, strong ML model-based identification does not by itself imply
open-set rejection capability. These distinctions bound the security claim and
motivate explicit and known--unknown evaluation in future
work.

%\textit{Remark:} Coordinate canonicalization does not eliminate sensor heterogeneity. Studies of smartphone IMUs report model-dependent bias, scale factor, noise, bias instability, random walk, and platform-processing differences \cite{kos2016crossplatform,kos2016biofeedback,capuano2023smartphone}. In a biometric setting, stable hardware characteristics can become an unintended identity cue when participants repeatedly use the same handset. This motivates both explicit device-confound discussion and future crossed participant--device validation.

\section{Conclusion}
This study shows that a prescribed prayer sequence contains measurable
participant-discriminative motion structure. Across 95 participants, complete
two-\rakaah{} units support strong identification and verification, while
complete-session holdout demonstrates that a substantial component persists
across independently acquired recordings. Two complementary IMU representation
strategies reach this conclusion under different assumptions. Their matched performance indicates that directional
Qibla-referenced canonicalization is useful but not universally superior to rotation invariant representation.

Signal-, descriptor-, and prayer-component analyses further localize identity
information to acceleration/gravity-relative dynamics, compact temporal and
spectral features, and selected bowing/prostration intervals. The
Qibla-referenced directional ablation shows that vertical and acceleration
components are particularly informative but that the strongest performance
requires complementary motion across all physical axes. Supervised and
SSL models confirm that participant structure can also be
learned directly from inertial windows. 

%a targeted Qibla-SimCLR extension further strengthens unit-level participant-template verification. 

Several limitations remain. First, there is a need to test the proposed method on heterogeneous hardware: session-disjoint
performance is not yet equivalent to device-disjoint behavioral validation.
Second, complete-session evaluation is restricted to participants with repeated recordings over a cumulative period of a few weeks, and therefore does not establish long-term biometric permanence over months or years. Third,
the data were collected during congregational Taraweeh prayer; individual
prayer may alter pacing and pause duration. Fourth, phone placement, clothing,
fatigue, temporary pain, and mobility limitations may perturb genuine-user
scores. Finally, the evaluation emphasizes ML model-based identification and
known-participant verification. Open-set rejection, targeted imitation, replay,
and sensor-level spoofing remain outside the present threat model. Future work will look into all aforementioned issues. This is necessary before the reported results can be interpreted as a deployment-ready authentication system.

%The targeted Qibla-referenced SimCLR extension is evaluated only under the unit-level protocol, so complete-session learned-model comparisons between the two coordinate representations remain outside the present evidence.

%The principal remaining scientific boundary is heterogeneous hardware: session-disjoint performance is not yet equivalent to device-disjoint behavioral validation. Separating human movement signatures from persistent smartphone characteristics, together with longitudinal and attack-aware evaluation, is therefore central to the next stage of this work.

%Future validation should therefore combine repeated longitudinal acquisition, cross-device recording, controlled placement variation, individual versus congregational prayer, and explicit open-set and attack-aware evaluation. These extensions are necessary before the reported results can be interpreted as a deployment-ready authentication system.

%LIMU-BERT and large-scale wearable studies further demonstrate that unlabeled motion data can support transferable representations \cite{xu2021limubert,yuan2022selfsupervised,ek2024selfsupervisedcomparison}.

\balance
\bibliographystyle{IEEEtran}
\bibliography{references}

%\clearpage
%\onecolumn
%\appendices

\clearpage
\onecolumn
\appendices
\section{Handcrafted Feature Definitions}
\label{app:feature_definitions}

Let
\begin{equation}
\mathbf{x}=[x_1,x_2,\ldots,x_N]^{\top}
\end{equation}
denote the samples of one derived signal within a 5-s window. Let $\mu$ and
$\sigma$ denote its population mean and standard deviation, $Q_p(\mathbf x)$
its $p$th quantile, and $\Delta x_n=x_n-x_{n-1}$ its first difference. For the
frequency-domain descriptors, the signal mean is removed before computing the
one-sided Fourier power spectrum and the zero-frequency component is excluded.

\begin{table}[!ht]
\centering
\caption{Definitions of the 20 descriptors computed for each derived
signal family. Applying the descriptors to five invariant signal families
produces 100 features per window; applying them to the six Qibla-referenced
channels produces 120.}
\label{tab:handcrafted_feature_definitions}
\renewcommand{\arraystretch}{1.02}
\setlength{\tabcolsep}{5pt}
\footnotesize
\begin{tabular}{cllp{7.1cm}}
\toprule
\textbf{No.}
&
\textbf{Descriptor}
&
\textbf{Category}
&
\textbf{Definition}
\\
\midrule

1 & Mean & Central tendency
& $\displaystyle \mu=\frac{1}{N}\sum_{n=1}^{N}x_n$ \\

2 & Standard deviation & Variability
& $\displaystyle \sigma=\sqrt{\frac{1}{N}\sum_{n=1}^{N}(x_n-\mu)^2}$ \\

3 & Median & Central tendency
& $\displaystyle Q_{0.50}(\mathbf{x})$ \\

4 & Minimum & Range
& $\displaystyle \min_n x_n$ \\

5 & Maximum & Range
& $\displaystyle \max_n x_n$ \\

6 & Range & Range
& $\displaystyle \max_n x_n-\min_n x_n$ \\

7 & Root mean square & Amplitude
& $\displaystyle
\mathrm{RMS}=\sqrt{\frac{1}{N}\sum_{n=1}^{N}x_n^2}$ \\

8 & First quartile & Distribution
& $\displaystyle Q_{0.25}(\mathbf{x})$ \\

9 & Third quartile & Distribution
& $\displaystyle Q_{0.75}(\mathbf{x})$ \\

10 & Interquartile range & Variability
& $\displaystyle
\mathrm{IQR}=Q_{0.75}(\mathbf{x})-Q_{0.25}(\mathbf{x})$ \\

11 & 5th percentile & Distribution
& $\displaystyle Q_{0.05}(\mathbf{x})$ \\

12 & 95th percentile & Distribution
& $\displaystyle Q_{0.95}(\mathbf{x})$ \\

13 & Median absolute deviation & Robust variability
& $\displaystyle
\mathrm{MAD}
=
\operatorname{median}_{n}
\left|x_n-Q_{0.50}(\mathbf{x})\right|$ \\

14 & Mean-square energy & Amplitude
& $\displaystyle
E=\frac{1}{N}\sum_{n=1}^{N}x_n^2$ \\

15 & Skewness & Distribution shape
& $\displaystyle
\gamma_1
=
\frac{1}{N}
\sum_{n=1}^{N}
\left(\frac{x_n-\mu}{\sigma}\right)^3$ \\

16 & Excess kurtosis & Distribution shape
& $\displaystyle
\gamma_2
=
\frac{1}{N}
\sum_{n=1}^{N}
\left(\frac{x_n-\mu}{\sigma}\right)^4-3$ \\

17 & Mean absolute first difference & Motion dynamics
& $\displaystyle
D_{\mathrm{abs}}
=
\frac{1}{N-1}
\sum_{n=2}^{N}
|\Delta x_n|$ \\

18 & First-difference RMS & Motion dynamics
& $\displaystyle
D_{\mathrm{rms}}
=
\sqrt{
\frac{1}{N-1}
\sum_{n=2}^{N}
(\Delta x_n)^2
}$ \\

19 & Dominant frequency & Frequency domain
& Frequency associated with the maximum nonzero-frequency power:
$\displaystyle f_{\mathrm{dom}}=f_{\arg\max_k P_k}$ \\

20 & Spectral entropy & Frequency domain
& $\displaystyle
H_{\mathrm{spec}}
=
-\frac{\sum_{k=1}^{K}p_k\log_2(p_k+\epsilon)}
{\log_2 K},
\quad
p_k=\frac{P_k}{\sum_{j=1}^{K}P_j}$ \\

\bottomrule
\end{tabular}
\end{table}

\end{document}